\documentclass[11pt,british]{article}
\usepackage[T1]{fontenc}
\usepackage[utf8]{inputenc}
\usepackage{geometry}
\usepackage{booktabs}
\usepackage{siunitx}
\newcolumntype{d}{S[
    input-open-uncertainty=,
    input-close-uncertainty=,
    parse-numbers = false,
    table-align-text-pre=false,
    table-align-text-post=false
 ]}
\usepackage{rotating}  

\usepackage{float}
\usepackage{makecell}
\usepackage{subcaption}
\usepackage{tabularray}
\usepackage{textcomp}
\usepackage{amstext}
\usepackage{setspace}
\usepackage[authoryear]{natbib}
\usepackage{hyperref}
\usepackage{color}

\makeatletter

\usepackage{authblk}

\usepackage{adjustbox}
\usepackage{graphicx}
\usepackage{array}
\usepackage{rotating}
\usepackage{rotfloat}
\usepackage{float}
\usepackage{booktabs}
\usepackage{stackrel}
\usepackage[skip=1pt, font={sc}]{caption}
\usepackage{varwidth}

\usepackage{amsmath}
\usepackage{amsfonts}

\usepackage{tikz}

\usepackage{placeins}
\usepackage{caption}
\renewcommand{\thetable}{\textbf{Table \arabic{table}}}

\defcitealias{CORPAUTH}{RLCAN}

\makeatother

\usepackage{babel}

\usepackage{array}
\newcolumntype{H}{>{\setbox0=\hbox\bgroup}c<{\egroup}@{}}

\begin{document}

\title{The temporary impact of permanent employment incentives: Evidence from Italy}

 \author[1,2]{Michele Cantarella}
 \author[2,3]{Maria Cristina Maurizio\thanks{Corresponding author. Email: \url{mariacristina.maurizio@imtlucca.it}}}
 \author[2]{Francesco Serti}
 \affil[1]{Technical University of Denmark - DTU}
\affil[2]{IMT School for Advanced Studies Lucca}
\affil[3]{University of Alicante}
\maketitle
\vspace{-1cm}
\begin{abstract}


This paper evaluates the short and medium-term effectiveness of hiring incentives aimed at promoting the permanent conversion of temporary contracts through social contribution exemptions. Using rich administrative data from Tuscany, providing detailed employment histories, we use difference in differences and regression discontinuity designs to exploit a unique change in eligibility criteria in 2018. We find that the incentives immediately increased the probability of conversion, with no evidence of substitution against non-eligible cohorts. However, these positive effects were short-lived and appear to reflect anticipated conversions, as we find null longer-term effects on permanent hirings.  



\end{abstract}

\newpage
\section{Introduction}

From the last years of the 1990s to the very beginning of the 2000s, most European countries sought to enhance labor market flexibility by liberalizing temporary contracts. These regulatory changes have facilitated worker flows by stimulating both hiring and separations. A key consequence in many countries has been the sharp expansion of fixed-term contracts, which has contributed to the emergence of dual labor markets, where short-term and temporary arrangements have come to rival and in some cases surpass open-ended contracts \citep{europeancommission2010employment}. Empirical evidence also suggests that under certain contractual arrangements or within specific institutional settings, fixed-term employment can be associated with fragmented career trajectories, lower earnings stability, and thus increased exposure to economic insecurity and social exclusion \citep[see, among others,][]{berton2011temporary, cappellari2012temporary,hijzen2017impact, GarcaPrez2018, Filomena2022}. 

To mitigate these potential negative consequences without reversing the liberalization of temporary contracts, policymakers of advanced economies, like Italy, Spain, and France,\footnote{Among others: for Spain law 43/2006, \textit{Tarifa Joven} (2014-2016); for \textit{France Contrat nouvelle embauche (CNE)} – 2005, Régime des emplois francs – 2020. See Section \ref{s:background} for further details on the Italian incentives.} have frequently implemented incentives to encourage firms to increase job stability. These policies typically involve bonuses, often in the form of payroll tax reductions or cuts in social security contributions, awarded to firms that either hire new employees under permanent contracts or convert temporary contracts into permanent ones. 

Our study focuses on the social contribution exemption introduced in Italy during the mid-2010s for employers hiring young workers into new open-ended contracts (OECs) or converting existing fixed-term contracts to permanent positions. Over the years, the scheme has undergone several changes to its eligibility criteria.\footnote{A detailed discussion is presented in Section \ref{s:background}.}
In particular, we leverage the unique exogenous discontinuity in treatment eligibility introduced by the 2018 Budget Law to estimate the causal impact of incentives on contract conversions, utilizing administrative worker-level data for Tuscany. With respect to the existing literature, we concentrate on permanent contract conversions rather than new hires, allowing us to study the effects of these incentives over a target population of workers not detached from the labor market, and to avoid confounding these effects with those of other concurrent incentives for temporary hires. We provide a comprehensive assessment of the reform across four key dimensions: the immediate effect on contract conversions, the medium-term impact on beneficiaries' career trajectories, the heterogeneous effects of the policy, and the substitution patterns between eligible and ineligible cohorts. 



The literature on the short-term effects of hiring subsidies often documents moderately positive results on employment among eligible workers during the subsidy period. For example, a generous one-shot wage subsidy in Belgium for unemployed youth (“Win-Win Plan” after the Great Recession) raised the likelihood of finding private-sector employment by about 10 percentage points in the first year for eligible young people \citep{Albanese2024}. In France, a one-year emergency hiring credit targeting low-wage jobs during the Great Recession led to significant employment gains \citep{cahuc2019effectiveness}, with no measurable increase in wages. 
Similarly, a large payroll tax cut for young workers in Sweden (2007–2009) led to a 2–3 percentage point increase in youth employment for the targeted age group, with treated firms hiring more young workers and expanding their overall workforce \citep{Saez2019}. 
All these incentives targeted first-time or disadvantaged low-paid workers, encompassing both permanent and temporary contracts.
These groups of beneficiaries may differ significantly from the general working population, limiting the relevance of their results in contexts where workers are already in temporary positions. 

By contrast, the hiring incentives introduced in Italy over the past decade have been designed primarily to promote more stable forms of employment within the general workforce. The literature, so far, has focused on the evaluation of the 2012 \citep{ciani2015getting} and 2015 \citep[see, among others,][]{sestito2018firing, boeri2019,deidda2021counterfactual,brunetti2022evaluating,ardito2023combined,Santoni2025} schemes, generally reporting a positive impact. However, these incentives often overlapped with broader schemes for younger workers that also covered temporary contracts, making it difficult to isolate the specific contribution of permanent hiring subsidies. The 2018 reform, by introducing an age threshold that extended eligibility to a group otherwise excluded from any other incentives, provides a unique opportunity to disentangle this effect.

In addition to these immediate impacts, some literature has also focused on the longer-term persistence of employment gains from hiring incentives, often in terms of job stability and career prospects. This evidence is mixed and not very encouraging. In the context of the aforementioned Belgian youth subsidy, \citet{Albanese2024} found that the positive impact for high school dropouts did not persist beyond the subsidy period, while high school graduates in the same program experienced some modest yet persistent career benefits. Focusing instead on the abolition of another Belgian hiring subsidy targeted at older long-term unemployed, \citet{desiere2022effective} also found that incentives mainly create temporary, short-lived employment. In contrast, \citet{Saez2021} studied the longer-term impacts of the 2007 incentives in Sweden, finding significantly positive employment effects for the subsequent career of the beneficiaries, a result that contrasts with the findings of \citet{egebark2018payroll}, who found little evidence of lasting effects for the treated individuals once they aged out of eligibility. Moreover, \citet{Sjgren2015}, analyzing another Swedish policy, and \citet{Batut2021}, examining a hiring credit for small firms in France, found that although employment declines once subsidies expire, workers who obtained subsidized jobs still exhibit a higher probability of being employed even after the expiration of the subsidies. Narrowing the focus to the 2015 Italian incentives, \citet{ardito2025effect} found that although the policy initially reduced separation risks, this effect was short-lived, with a sharp spike in exits occurring precisely when the subsidy expired. As with the short-term evidence, these contrasting results may arise from the presence of overlapping schemes, the focus of incentives on specific populations, or both.


Other strands of literature have focused on spillover and substitution effects. A key concern is that subsidizing some jobs or workers may simply displace employment opportunities elsewhere. The available evidence, primarily concerning first-time labor market entrants, generally indicates negative or negligible spillovers on non-eligible workers. For example, the French low-wage hiring credit of 2009 showed no displacement of incumbent workers or non-eligible hires at treated firms \citep{cahuc2019effectiveness}. Likewise, evaluations of youth-focused subsidies in Belgium detected minimal substitution away from slightly older (ineligible) workers \citep{Albanese2024}. Similarly, \citet{ciani2015getting}, analyzing the 2012 Italian incentives, found no substitution effects. The same limitations noted earlier also restrict the generalizability of these studies to the effects on open-ended contract conversions. 

In short, a comprehensive evaluation of hiring incentives that focuses on the transition from temporary to permanent contracts, and that disentangles these incentives from more general (applicable to any contract) and targeted (applicable to specific vulnerable groups) employment incentive schemes, remains missing from the literature. 
In our study, we identify the policy effects of the 2018 Italian incentives on contract transitions by using a combination of difference-in-differences and regression discontinuity designs. We find that the incentives effectively increased the immediate conversion rate from temporary to open-ended contracts for eligible workers. The effect is more pronounced for men, and for those working in the industrial and service sectors, in medium-skilled occupations, and in small firms. To account for overall substitution effects with slightly older peers, we perform placebo tests in the treatment year, obtaining null results. We also find no within-firm substitution effects as the change in the proportion of eligible workers around the cutoff with respect to the previous year does not appear to significantly affect the conversion rate of eligible workers. However, we do not detect any significant longer-term effect one, two, three, and four years after the introduction of the policy; indeed, the incentives' effects, and their heterogeneities, disappear within a year of the reform. This evidence suggests that the policy merely anticipated the conversion to a permanent contract for temporary workers who would have been converted anyway. 

This paper is structured as follows. Section \ref{s:background} provides the reader with a background of incentive schemes in Italy. Section \ref{s:data} details our data sources, while Section \ref{s:method} discusses our identification strategy. Our results are presented in Section \ref{s:results}, and Section \ref{s:conclusions} concludes.

\section{Institutional Background}
\label{s:background}

Following the financial and sovereign debt crisis, social contribution cuts for open-ended hires in Italy have been used since the mid-2010s as a way to encourage conversions from temporary to permanent employment for young workers. The eligibility criteria and generosity were changed repeatedly by successive budget laws and reforms.

The first set of these incentives was introduced as part of the 2012 \textit{Monti-Fornero reform} (Law 92/2012), among other labor market reforms meant to regulate temporary contracts and reduce dismissal costs under specific circumstances.
The incentives initially encompassed the hiring, under both permanent and temporary contracts, of people over 50 years old and disadvantaged women.\footnote{Were included: women of any age, residing anywhere, who have been without regular paid employment for at least 24 months or if they were living in disadvantaged areas or working in sectors with high gender employment disparities, the limit was lowered to 6 months.} The incentives scheme entitled the employer to a 50\% reduction in the social contribution payment for the hiring of eligible workers for up to 18 months, provided that it results in a net increase in the company’s workforce. These incentives, which still remain in effect, were accompanied by an extraordinary incentive scheme\footnote{Decree October 5th 2012} which included men below 30 years old and women of all ages, and granted employers an exemption ranging from EUR 3,000 for short temporary contracts to EUR 12,000 for conversion toward permanent contracts. However, given the limited amount of available funds, this subsidy was depleted in just a few days. 
\citet{ciani2015getting} examined the immediate effects of the extraordinary scheme by focusing on its effect on temporary-to-permanent transitions. 

Subsequent programs targeted youth unemployment more explicitly. Following a recommendation of the European Council,\footnote{COUNCIL RECOMMENDATION of 22 April 2013 on establishing a Youth Guarantee (2013/C 120/01); Available at: \url{https://eur-lex.europa.eu/LexUriServ/LexUriServ.do?uri=OJ:C:2013:120:0001:0006:EN:PDF}.} the Youth Guarantee Program (\textit{Programma Nazionale Garanzia Giovani}, YG henceforth) was established in Italy in 2014. The set of laws included incentives for employers to hire young people on permanent, fixed-term, or apprenticeship contracts, ranging from 1,500 euros to 6,000 euros depending on the type of contract and the employability of the individual. This program has survived, with some changes, until 2020, and subsequently has been included in the 2021-2027 EU-wide program Youth, Women and Work (\textit{Programma Nazionale Giovani, Donne e Lavoro}). 

The biennium 2014–2015 marked a period of significant reforms in Italy, many of which were included in the package known as the \textit{Jobs Act}. 
The major changes sought to incentivize hiring by making the Italian labor market more flexible, and entailed the reduction of firing costs for workers with OECs, which was complemented by the introduction of generous hiring incentives. This incentive scheme consisted of a three-year exemption from the full amount of social security contributions for all open-ended contract hires and transition from fixed-term ones made during that year, with no age restrictions, for a maximum of EUR 8,000. The entanglement between all these measures, including the incentives included in the Youth Guarantee, has complicated the evaluation of the Jobs Act. Nevertheless, the studies focusing on such evaluation \citep{sestito2018firing,boeri2019,deidda2021counterfactual,brunetti2022evaluating,ardito2023combined,Santoni2025} found small to moderate effects, mostly attributable to the incentives rather than to the reduction of firing costs.
Initially, this incentive was intended to cover both 2015 and 2016, but with the 2016 Budget Law (L. 208/2015), the exemption from social security contributions payment was changed to a 40\% reduction for the hirings and transitions in 2016, for a maximum duration of 2 years. 

With the 2017 Budget Law and other minor decrees, hiring incentives were reintroduced through several fragmented schemes. The most relevant for our context is that the incentives under the YG program were significantly upgraded. Encompassing all under-30 hires, the updated deductions amounted to 100\% in the case of new permanent contracts (including conversions) and 50\% in the case of temporary contracts, with a maximum duration of one year. Other concurrent measures targeted specific regions (Occupazione Sud) or recent high-school graduates with school-to-work experience at the same firm. None of these measures applied to the population we focus on in this paper (i.e., Tuscan workers in their 30s).

Finally, the 2018 budget law (Law n. 205/2017) updated the employment incentives for new and converted permanent hires, keeping the incentives for temporary contracts unchanged. While the generosity of the former incentives was reduced, entailing a 50\% reduction in social contributions (up to a maximum annual amount of EUR 3,000), the duration was increased to 36 months, and the age limit threshold was raised to 35 years old. 
This change in cohort eligibility was intended to be a temporary one, and was meant to return to 30 for the subsequent years, 2019 and 2020. However, at the end of 2019, with the new Budget Law for 2020 (Law n.160/2019), the threshold was retroactively set back to 35 for 2019 and then maintained for 2020 as well. While technically, employers were not aware of the retroactive eligibility change until the end of that year, it is unclear if they were actually able to anticipate it.

2018 was also marked by a significant labor market reform. In July, the so-called \textit{Decreto Dignità} was approved, marking a partial reversal of the Jobs Act and responding to a Constitutional Court ruling that had struck down parts of its dismissal rules. 
Although this policy did not directly affect the incentives under study, leaving the cutoff and the generosity of the incentives unchanged, tighter regulations on temporary contracts may have increased the likelihood of permanent conversions, especially for workers already close to the new maximum duration. However, the reform only became completely effective in November 2018, allowing us to estimate the immediate effect without major drawbacks. Furthermore, since the 35-year age threshold was not affected by the reform, workers around the cutoff were treated uniformly, thereby maintaining the internal validity of our results. This being considered, in Appendix \ref{s:ddignita}, we show how the estimated short-term policy effect is unaffected by the introduction of this concurrent policy.


An additional discussion concerns the implementation of these schemes and their restrictions. With the exception of the 2012 extraordinary incentives, the implementation of the incentives has been almost automatic for employers. In practice, firms applied the exemptions directly when calculating their social security contributions and reported them through the standard INPS system, later consolidated in their annual tax filings. An additional restriction was introduced in 2017: employers could not benefit from hiring incentives if they had carried out dismissals in the same production unit within the preceding six months. This “anti-layoff” clause was meant to prevent abuses in which subsidized youth hires were immediately offset by terminations of existing staff.

\begin{figure}
    \centering
    \includegraphics[width=0.8\linewidth]{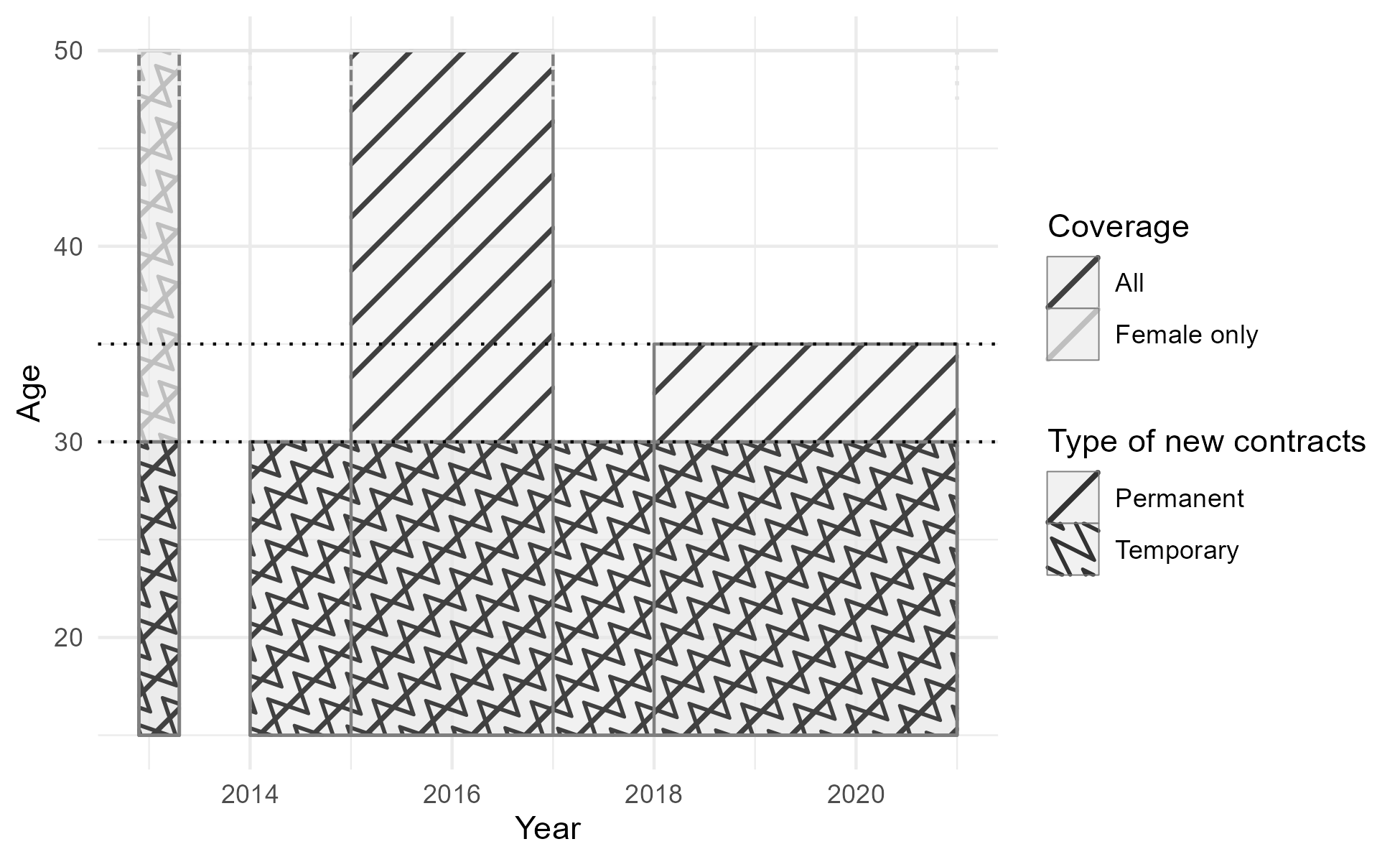}
    \caption{\textsc{Employment incentives over the decade} \\ Eligibility window, along with age and type of contact eligibility for each incentive scheme.}
    \label{fig:timeline}
\end{figure}

\hyperref[fig:timeline]{Figure~\ref*{fig:timeline}} summarises the evolution of the incentive schemes over the decade. The figure highlights that the 2018 change in the age eligibility threshold is particularly important: for the first time, it created a sharp and unambiguous distinction between eligible and ineligible cohorts for permanent hirings. This discontinuity provides clear leverage for causal identification, in contrast to earlier reforms where overlapping policies made it difficult to disentangle the effects of individual measures.

\FloatBarrier

\section{Data}
\label{s:data}

Our study was carried out using data from Tuscany, the fifth most-populated Italian region.\footnote{Source: $https://www.istat.it/wp-content/uploads/2024/05/Toscana\_Focus2022.pdf$} The data originate from the labor information system of the regional government, where, since March 2008, employers have been required to report new hires or changes to existing contracts via an online platform. This set of information is known as the system of mandatory communication (\textit{Sistema di Comunicazioni Obbligatorie}). 

These communications fall into four categories: hiring, extensions (\textit{proroghe}), conversions (\textit{trasformazioni}),\footnote{I.e, tenure changes from temporary to permanent.} and terminations (\textit{cessazioni}). A detailed set of worker and job-specific characteristics is also included with each communication. Together, these mandatory communications allow us to reconstruct labor flows that have occurred since the system was implemented, covering around 12 million job spells in Tuscany alone. The dataset links job contracts with unique worker and firm identifiers, allowing us to track the evolution and, possibly, conversion of temporary contracts over time.


To best study conversions, we start by considering the set of mandatory communications concerning standard temporary contracts\footnote{Standard temporary contracts, vis-a-vis other atypical ones, represent the most common form of fixed-term contract in the Italian labour market. In fact, according to a report from the National Social Security Institution (INPS), standard temporary contracts represented 45\% of the total number of new hires in 2018, while atypical contracts accounted for only 33\%. Accordingly, atypical contracts are not considered in our analysis. Source: \url{https://www.inps.it/content/dam/inps-site/pdf/dati-analisi-bilanci/osservatori-statistici/osservatorio-precariato/5328KEY-osservatorio_precariato_gen_dic_2018.pdf?utm_source=chatgpt.com}} for each year of our period of interest, ranging from 2014 to 2019. For each calendar year, we select all the job relationships on standard fixed-term contracts held by individuals aged 34–36 who have never previously held a permanent contract. In the next section, we look at a wider group of individuals in order to present some preliminary evidence.




\subsection{Descriptive evidence on policy take-up for conversions}
\label{s:descriptive}

\begin{figure}[!ht]
    \centering
\includegraphics[width=0.8\linewidth]{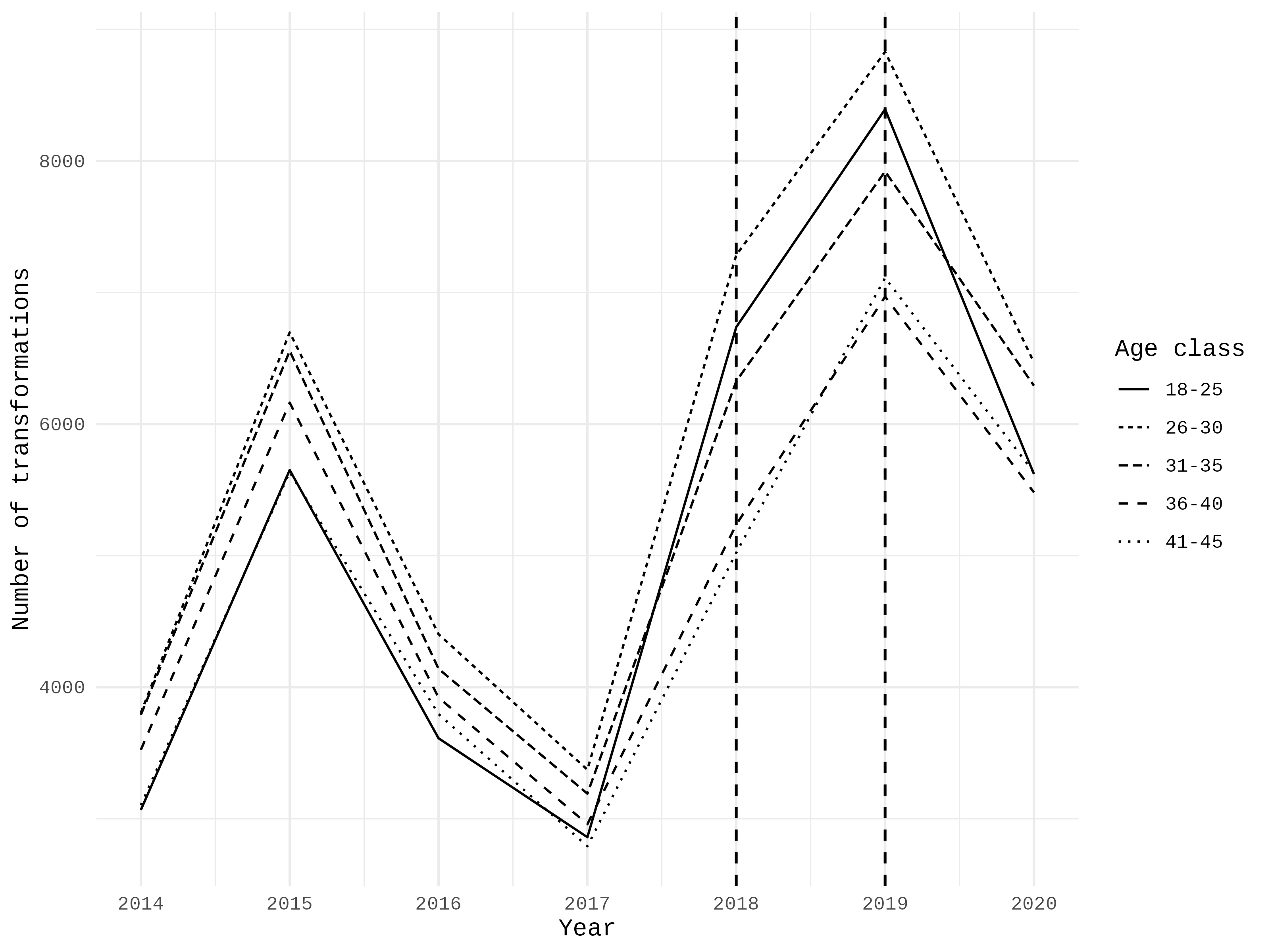}
\caption{\textsc{Conversions by age-groups} \\ The plot depicts the number of conversions of fixed-term contracts to open-ended contracts, grouped by 5-year age classes. The vertical lines represent the policy window during which the age limit was originally set to be 35.}
    \label{fig:fig1}
\end{figure}

\hyperref[fig:fig1]{Figure~\ref*{fig:fig1}} plots the absolute number of conversions by 5-year age cohorts, providing initial exploratory evidence on the policy take-up effect over the years.

Several stylized facts are immediately evident from the figure. Firstly, both the 2015 and 2018–2019 windows are characterised by significant overall increases in conversions. Nonetheless, the trends in conversions appear to be parallel among the age groups. A single discontinuity emerges among the cohorts, appearing only in 2018, further motivating our analysis. In that year, conversions rose across all age groups, but the increase was noticeably smaller for those aged 35 and older. Interestingly, the reintroduction of hiring incentives in 2017 was not accompanied by any differential increase in conversions for the eligible population of under-30s. Possibly, the combination of incentives for temporary and permanent hires for the same age group was detrimental to permanent conversions.

\begin{figure}[!ht]
    \centering
\includegraphics[width=0.8\linewidth]{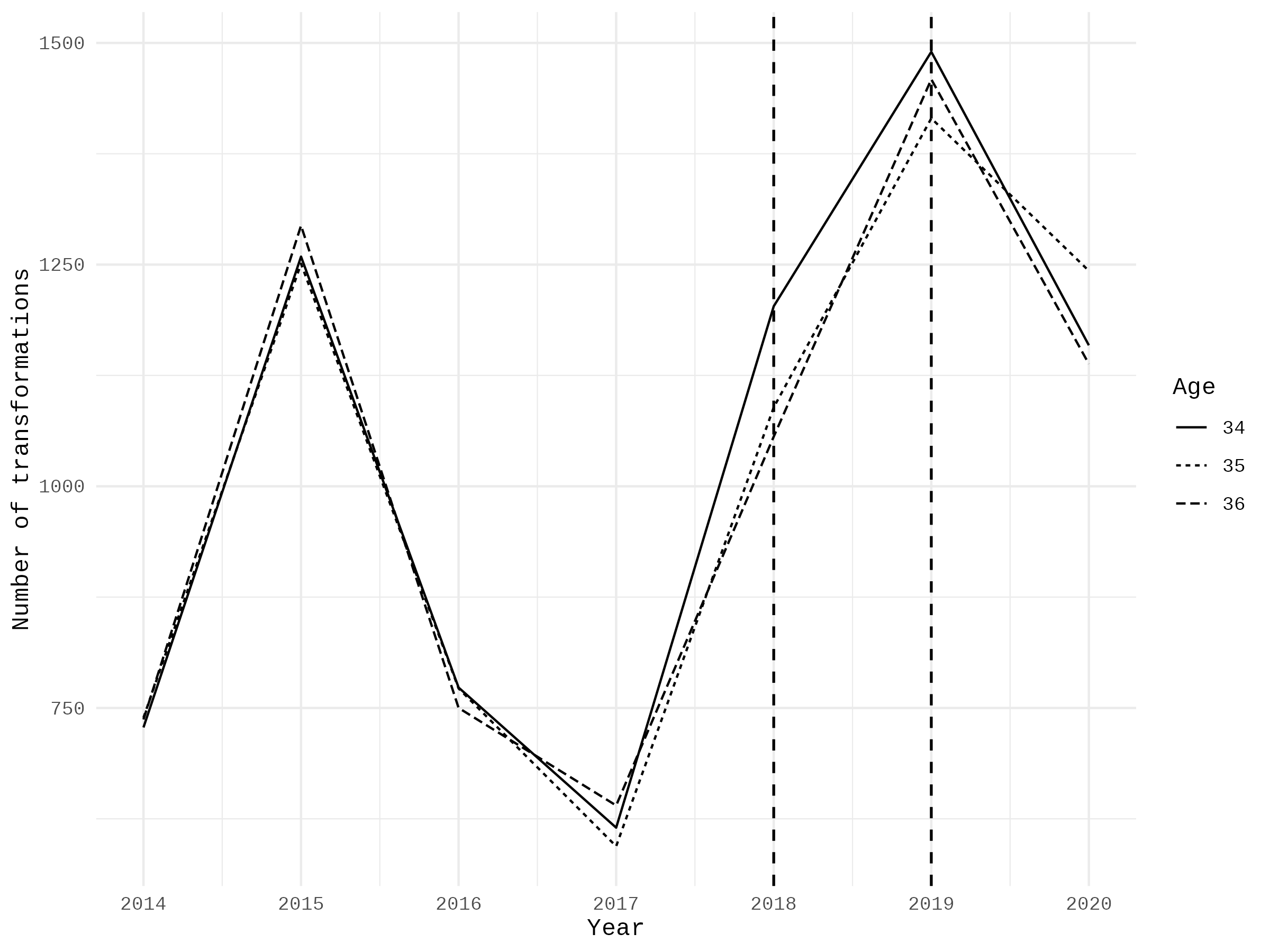}
    \caption{\textsc{Conversions by age close to the policy threshold} \\In the plot, the number of conversions of fixed-term contracts to open-ended contracts is depicted, focusing on the one for individuals aged between 34 and 36. The vertical lines represent the beginning and the end of the period during which the age limit was known to be 35.}
    \label{fig:fig2}
\end{figure}

\hyperref[fig:fig2]{Figure~\ref*{fig:fig2}} further disentangles these trends by looking at individuals turning 34, 35, and 36 in each year. Even after these sample restrictions, the gap between the 34- and 35-year-old groups increased significantly in 2018. The conversion for people aged 36 years old in 2018 increased, but at a lower pace. The previously mentioned gap narrowed significantly in 2019, which aligns with the fact that these individuals were initially excluded from coverage that year.. 

While these descriptive patterns are suggestive, they may also reflect other confounding factors. In the next sections, we formalize these intuitions in a DiD framework, and then sharpen the focus by exploiting variation around the birthday cutoff in a regression discontinuity design.

\FloatBarrier

\section{Econometric models}
\label{s:method}

\subsection{Difference in differences}


Using yearly repeated cross-sections, we first employ a difference-in-differences (DiD) approach, exploiting the introduction of the 35-year-old eligibility threshold in 2018. Accordingly, each year we define the treatment and control groups based on the age of the employee, and define the 2018-19 window as the treatment period. 
We include in our sample all temporary contracts that were active for at least one day between 2014 and 2019, and held, each year, by individuals aged between 33 and 36 who had never held a permanent contract before.

Since the benefits are applied automatically when the firm files the mandatory social contributions form, our data allows us to unambiguously identify eligibility for the incentives. Therefore, by comparing eligible and non-eligible cohorts, we aim to identify the Intention to Treat (ITT).



The DiD method is one of the most widely used econometric tools adopted in policy evaluation studies \citep{CALLAWAY2021200, 10.1093/ectj/utad016}.  
The main identifying assumption of the DiD is that trends across eligible and not eligible cohorts would have kept evolving in parallel in the absence of the policy (i.e., the parallel trends assumption). However, our repeated cross-sectional setting implies that observations across time periods are not drawn from the same underlying population, imposing an additional \textit{no-compositional change assumption} \citep{sant2023difference}. This additional assumption entails that samples, though potentially composed of different individuals, should exhibit comparable distributions of observable characteristics. In our case, it seems natural for this condition to hold given that the treatment is only related to the age of the employee, meaning that imbalances of group composition should not be attributed to endogenous worker characteristics. We have nonetheless tested this additional requirement by comparing the distribution of the principal covariates in the annual samples, finding almost no significant compositional differences in observable covariates between groups. The results of this test are available in Appendix \ref{Appendix:covariates}, Table \ref{tab:balance}.

In our practical implementation, we adopt an Event Study Difference-in-Differences design, estimating a separate treatment effect intercept for each year in the sample.
The estimating equation is the following: 
\begin{align}
    y_{it} = \alpha + D_{i} \theta_1  +\sum_{t=-4}^{1}R_{t}D_{i}\theta_{1,t} +\sum_{t=-4}^{1}R_{t}\delta_t+  X_{i}^{'}\beta + u_{it}
\end{align}

where $y$ is the short and longer-term outcome of interest for individual $i$ observed at time $t$. We detail the analyzed outcomes below in subsection \ref{sub:outcome}. The treatment group indicator is denoted by $D_i$, indicating age eligibility, while time intercepts are denoted by $R^k_{i}$, with $k$ denoting the lags and leads from the 2018 treatment period. The interaction of the treatment with the time variable in 2018 identifies the ITT (\textit{Intention to treat}) effect $\theta_{1,0}$. 

We implement two different specifications for $D_i$. The first specification uses a standard binary treatment variable, which applies only to individuals turning 34 or 36, leaving 35-year-olds out of the sample. The variable equals 1 for the eligible age group and 0 for the ineligible one. 
The second specification follows the approach of \cite{cheng2013does}, defining the treatment intensity of 35-year-olds (now re-included in the sample) as the share of the year spent at age 34; that is, the proportion of the year in which they were potentially eligible for the incentives. To construct this variable, we account for whether an individual’s birthday occurred before or after the 15th of the month, assigning treatment status at the level of whole months accordingly.


Finally, $X_{i}$ denotes the employee and contract characteristics. Specifically, we include gender, education level (divided between "Elementary school", "Middle school", "High school", "Degree and more", and "No level reported"), and a dummy variable for Italian citizenship for our \textit{individual-level variables}. \textit{Job-level information} includes the industrial sector (using the \textit{ATECO} 1-digit classifications), the skill level (using the \textit{Classificazione delle professioni 2011}\footnote{The \textit{Classificazione delle professioni 2011} is the Italian version of the ISCO dataset https://www.istat.it/classificazione/classificazione-delle-professioni/} 1-digit classifications), and the place of work (11 levels, including all Tuscany provinces and an out-of-Tuscany class including all the observations with workplace outside of Tuscany). We also account for the starting month of the contract, to capture potential seasonal or cyclical patterns in the labor market, and the potential duration of the contract, computed as the difference between the end of the year and the contract’s starting date—since the actual duration cannot be used due to its endogeneity with respect to the outcome variable.

\subsection{Regression discontinuity design}

The DiD specifications discussed above rely on the assumption that age groups are inherently comparable over the years. We further relax this assumption in a Regression Discontinuity design (RD) by looking at changes in employment around the 35th birthday in the treatment year only, using daily data. Using distance from the 35th birthday in days as the running variable, we assess whether the probability of conversion changes discontinuously at the threshold.

For this analysis, we focus on people who turned 35 in 2018 and start by considering a symmetric time window of three months before and after the birthday. To ensure consistency, we restrict the sample to days within the calendar year 2018 in which the individual held an active temporary contract. As we observe employment outcomes for each individual and for every value of the running variable around the birthday, the setting resembles an event-study RD. 





The econometric specification is the following:
\begin{align}
    y_{id} = \alpha +  D_{d} \theta_1+ f(d) \theta_2  + f(d) D_{d} \theta_3  +  X_{i}^{'}\beta + u_{id}
\end{align}

where the variables are the same as in the binary DiD estimation, except for the running variable $d$, which represents the distance from the 35th birthday computed in days. The outcome variable now denotes the conversion outcome for individual $i$ at day $d$ (i.e, the \textit{change} from day to day). The treatment effect is now identified by $\theta_1$, identifying the jump in the probability of conversion at the age threshold.

To determine the appropriate functional form of $f(T_i)$, we conducted an exploratory analysis using polynomial specifications up to the fifth degree. As we observed that the function’s behavior stabilized after the second degree, we decided to retain only the linear and quadratic approximations in the initial estimation (based on the widest selected window) and restricted the final estimation to the linear specification, following \citep{Cattaneo2019}. As far as bandwidth around the threshold is concerned, we set a maximum symmetric window of three months, but then applied the \textit{Optimal Bandwidth Selection} method of \cite{Calonico2019}, to pick the most appropriate window. 

In terms of comparability with the DiD results, it is worth noting that the RD and DiD capture different parameters: the RD measures the local discontinuity in the daily probability of conversion at the eligibility threshold, whereas the DiD recovers the average annual effect for the eligible cohort. To facilitate comparison, RD estimates can be rescaled (e.g. to yearly probabilities) or expressed relative to the average conversion rate, but they should still be interpreted as local effects at the cutoff. Accordingly, we expect RD estimates to be larger: RD focuses on the salient, fully treated cutoff day, whereas DiD dilutes the effect by averaging over the year and across workers whose treatment is active only during days of fixed-term employment. 



\subsection{Difference in discontinuities}


As a final exercise, we then combine both RD and DiD approaches by comparing conversion rates around the age-cutoff in 2018 against the same effect in the other years in the sample, when the age eligibility threshold was either lower, retroactively implemented, or not applicable altogether. In the presence of additional unobserved policies or treatments that are active at the same age-cutoff of the policy we are interested in (i.e., the hiring incentives) both in 2018 and in the previous years (and under the assumption that the effects of these unobserved treatments are constant in time and do not interact with the effect of hiring incentives in 2018), this setup allows us to wash out these confounding treatments and more sharply identify the effects of hiring incentives of 2018.


We then employ a Difference in Discontinuities design \citep{Grembi2016}, known also as RD-DiD (henceforth). We use the same sample selection strategy as in the RD model discussed above, but extend the analysis to cover the years 2014–2017. The specification is the following:



\begin{multline}
    y_{idt} = \alpha + \sum_{t=-4}^{1}  R_{t}D_{d} \theta_{1,t} +    \sum_{t=-4}^{1}  R_{t}f(d) \theta_{2,t} +\sum_{t=-4}^{1}  R_{t}f(d)D_{d} \theta_{3,t} +\sum_{t=-4}^{1}R_{t}\delta_t+ \\ D_{d} \theta_1 + f(d)D_{d} \theta_2 + f(d) \theta_3  + X_{i}^{'}\beta + u_{idt}
\end{multline}

where we re-introduce the time indicators $K_i^k$. The indicators are interacted with the age-cutoff treatment, yielding the treatment effect $\theta_{1,0}$, denoting the discontinuity effect in the treatment year. In addition, the $K_i^k$ are interacted with the running variable, as well as jointly with the age-cutoff treatment and the running variable, to allow for differences in slopes across periods and treatment status.


\subsection{Short and longer-term effects}
\label{sub:outcome}

In evaluating the immediate take-up of the policy, our primary outcome of interest is a binary variable $y_{i}$, which takes the value of one if the job relationship $i$ transitions from temporary to permanent. It is set to 0 in all other cases, such as when the contract remains temporary and ends naturally or when it is terminated by either the employer or the employee. Consequently, the interpretation of the main results focuses on how the policy influences the likelihood of experiencing a conversion compared to any other possible outcome. 

For the longer-term medium run effects, it is crucial to examine whether the short-run impacts persist and genuinely translate into stable employment \citep{desiere2022effective, Saez2021}. While in our setting the conversion from fixed-term to permanent contracts might appear to guarantee permanence, it is important to note that permanent contracts can still be terminated. This issue has become even more salient following the introduction of the "contratto a tutele crescenti", a new form of open-ended contract introduced in Italy in 2015, which progressively increases employment protection with tenure and allows for easier dismissal in the early years of the relationship. $y_i$ denotes a set of a binary variables indicating whether the worker (i) still has the same permanent contract, (ii) still has a permanent contract (no matter where), (iii) still has a contract of any sorts with the same firm or in the same sector and (iv) the number of days worked $n$ years after the policy. We let $n$ take the values of 1, 2, 3,  and 4 years after the observed year. However, given the availability of our data, we had to drop the observations after the policy, as we are unable to follow the individuals selected in 2019 throughout the subsequent four years.

All models outlined above can be used to estimate the short-term effects. The longer-term effects, however, are only estimable through the difference-in-differences model. This is a well-known issue in dynamic RD settings where the same individual can be exposed to the same treatment multiple times \citep{Hsu2024}. In our RD design, identification relies on exploiting the smooth variation in outcomes within a narrow window around the 35th birthday. This is appropriate in the short term, since the exact timing of the birthday relative to the contract conversion window determines whether a worker is eligible or not.  In the longer term, however, outcomes such as remaining in the same permanent contract, keeping a permanent contract in general, or continuing employment in the same firm or sector become effectively invariant to the precise day on which the worker turned 35. After one, two, or three years, the running variable no longer provides meaningful within-window variation, and employment changes during the future time window are unlikely to be associated with the turning of age around the 35-year threshold. Put differently, individuals in our RD sample act as both treatment and control units, since the same worker appears on both sides of the cutoff. For longer-run outcomes, which are effectively day-invariant, the only meaningful comparison is between those who are always treated (e.g., 34-year-olds) and those who are never treated (e.g., 36-year-olds). In this sense, the RD (and RD-DiD) structure collapses into a DiD framework, where identification of longer-run effects comes from comparing eligible and non-eligible cohorts across periods.  

\subsection{Heterogeneous effects}

In addition to the overall policy effects, we consider additional specifications to study heterogeneous effects.  While some of the interacting dimensions may be endogenous to employment outcomes, identification relies on these characteristics being orthogonal to the implementation of incentives.
We consider the following effects:

\paragraph{Macrosector and firm size effects} We first analyze heterogeneous exposure to the policy across industries, examining whether take-up differed by sector, identified by three macrosectors: Agriculture (ATECO section A), Industry (ATECO sections C–F), and Services (ATECO sections G–U), as well as by firm size. Since our data do not provide direct information on firm size, we construct a proxy based on the number of hires recorded within each year.

\paragraph{Temporary contact length effects} Next, we examine heterogeneity by the start date of the temporary contract. In fact, contracts stipulated very close to the end of the year may be less likely to benefit from the incentive, partly because of possible bureaucratic delays, and partly because firms might prefer to convert workers who have spent more time with them. Furthermore, the date of hiring is also correlated with the individual's exposure to the policy (i.e., the number of days worked under a fixed-term contract). 


\paragraph{Demographics and job characteristics}  We also analyze heterogeneities by gender and skill level of the occupation, divided into low ("Classificazione delle professioni - CdP"\footnote{"Classificatione delle professioni" is the Italian version of ISCO} one digit 8-9), medium (CdP one digit level 4-7), and high (CdP one digit level 1-3). 


\subsection{Spillover effects} 

As an initial check for potential spillover effects, we re-estimate the DiD using alternative control groups, replacing the baseline controls with individuals aged 37, 38, or 39 in turn. This allows us to examine whether estimated policy effects differ when building the counterfactual by using the cohorts closer to the eligibility threshold, who may be more likely to be substituted by eligible workers. If substitution effects between eligible and non-eligible workers were indeed present, we would expect to observe a significantly smaller policy effect when using these additional cohorts as control groups, indicating that individuals closer to the threshold bore the negative consequences of increased conversions among the eligible. 

Then, we further analyze spillover effects by studying whether the proportion between eligible and non-eligible temporary workers (excluding worker $i$) influences the hiring of the former, under the assumption that the share of people above and below the age eligibility threshold is orthogonal to the policy.

In the DiD framework, we define this measure as the within-workplace proportion of eligible temporary employees aged 34 to 36 at the beginning of the year ($t-1$), which we interact the treatment and relative year dummies. A significant negative effect of the interacted term in the treatment year would indicate that eligible individuals were more likely to be hired when more of their peers were ineligible, suggesting a within-firm substitution effect. 

In the RD and RD-DiD models, for each worker $i$, we compute a weighted proportion of the younger ($D_k=1$) coworkers $k$ in the same estimation window, where the weight is inversely proportional to the age distance from worker $i$:

\begin{equation}
S_i = \frac{\sum_{k \neq i} D_k\frac{1}{1 +|T_i - T_k|}}
            {\sum_{k \neq i} \frac{1}{1+|T_i - T_k|}} ,
\end{equation}

Where $D_k$ indicates if $T_i \geq T_k$. Since not all firms employ workers in adjacent age groups, we restrict estimation of these spillover effects to the subsample of firms where workers of similar ages are present.

\section{Results}
\label{s:results}

\subsection{Short-term Results}
\subsubsection{Difference in Differences}
In order to evaluate the short-run effects of the policy, we begin by presenting the results from the standard binary Difference-in-Differences (DiD) regressions, estimated using individuals turning 34 or 36 in each year. These results are reported in the first three columns of Table \ref{tab:did_estimates}. The baseline specification (model 1) includes only the treatment indicator, the set of relative time dummies, and their interaction terms. The second specification (model 2) augments this baseline by introducing the set of covariates described in the previous section, while still assuming independent and identically distributed (IID) errors. The third specification (model 3) further extends the second one by clustering standard errors at the workplace level, thereby addressing potential within-workplace correlation in the error structure. 

Across all three specifications, the interaction between the treatment variable and the treatment period dummy (t = 0), corresponding to the year of policy implementation (2018), yields the coefficient of primary interest. The estimated effect is consistently positive, statistically significant, and remarkably stable in magnitude. The estimated effect ranges between 0.014 and 0.016 percentage points, which corresponds to an increase of approximately 32\%–36\% relative to the mean conversion rate. This finding suggests that firms actively responded to the introduction of the policy by increasing permanent conversions among eligible workers. Interestingly, the estimates also reveal a positive, albeit smaller, coefficient for the period immediately after implementation (t = 1), which may reflect firms’ expectations regarding the retroactive reinstatement of the 35-year-old threshold that would occur at the end of 2019. which applied to conversions carried out during that year (see section \ref{s:background}).

In columns 4 and 5, we re-include the 35-year olds in the sample and and turn to a continuous treatment denoting the proportion of days in 2018 over which individuals were eligibile to the policy. Here we replicate the first and third formulations of the binary DiD analysis, namely the baseline specification without covariates and the richer specification with covariates and clustered errors. The results we obtain are highly consistent across both definitions of the treatment variable. In particular, the estimated coefficients remain positive, statistically significant, and of comparable magnitude, which reinforces the robustness of the findings and suggests that the evaluation is not overly sensitive to the precise operationalization of the treatment. As in the binary case, the inclusion of covariates strengthens the estimated effect, both in size and statistical significance, further confirming the stability and reliability of the policy’s short-run impact. Furthermore, in the continuous specifications, the effects in 2019, which could have suggested an anticipatory effect of the policy threshold reinstatement, virtually disappear.

\begin{table}[h!]
\centering
\title{\textbf{TABLE 1: Difference in Differences estimation}}
\caption{}
\label{tab:did_estimates}
\resizebox{\textwidth}{!}{
\begin{tabular}{lccc|cc}
\toprule
\midrule
 & \multicolumn{3}{c}{\textbf{Binary treatment}} 
 & \multicolumn{2}{c}{\textbf{Continuous treatment}} \\
\cmidrule(lr){2-4} \cmidrule(lr){5-6}
 & Model 1 & Model 2 &  Model 3 
 & Model 4 & Model 5 \\
\midrule
Eligible × (t = 0)  & \num{0.012}** & \num{0.014}*** & \num{0.014}*** & \num{0.011}** & \num{0.013}*** \\
 & (\num{0.005}) & (\num{0.005}) & (\num{0.005}) & (\num{0.005}) & (\num{0.005}) \\
Eligible × (t = -4) & \num{0.001} & \num{0.005} & \num{0.005} & \num{0.002} & \num{0.005} \\
 & (\num{0.005}) & (\num{0.005}) & (\num{0.004}) & (\num{0.005}) & (\num{0.004}) \\
Eligible × (t = -3) & \num{-0.001} & \num{0.004} & \num{0.004} & \num{-0.001} & \num{0.003} \\
 & (\num{0.005}) & (\num{0.005}) & (\num{0.005}) & (\num{0.005}) & (\num{0.004}) \\
Eligible × (t = -2) & \num{0.002} & \num{0.006} & \num{0.006} & \num{0.002} & \num{0.006} \\
 & (\num{0.005}) & (\num{0.005}) & (\num{0.005}) & (\num{0.005}) & (\num{0.004}) \\
Eligible × (t = 1)  & \num{0.010}* & \num{0.013}*** & \num{0.013}** & \num{0.007} & \num{0.009}* \\
 & (\num{0.005}) & (\num{0.005}) & (\num{0.005}) & (\num{0.005}) & (\num{0.005}) \\
Eligible            & \num{0.003} & \num{-0.001} & \num{-0.001} & \num{0.003} & \num{-0.001} \\
 & (\num{0.004}) & (\num{0.004}) & (\num{0.003}) & (\num{0.003})  & (\num{0.003}) \\
t = -4               & \num{-0.006}* & \num{0.000} & \num{0.000} & \num{-0.008}** & \num{-0.002} \\
 & (\num{0.004}) & (\num{0.004}) & (\num{0.003}) & (\num{0.003}) & (\num{0.003}) \\
t = -3               & \num{0.014}*** & \num{0.020}*** & \num{0.020}*** & \num{0.013}*** & \num{0.020}*** \\
 & (\num{0.004}) & (\num{0.004}) & (\num{0.004}) & (\num{0.003}) & (\num{0.003}) \\
t = -2               & \num{0.007}* & \num{0.005} & \num{0.005} & \num{0.006}* & \num{0.005} \\
 & (\num{0.004}) & (\num{0.004}) & (\num{0.004}) & (\num{0.003}) & (\num{0.003}) \\
t = 0                & \num{0.019}*** & \num{0.012}*** & \num{0.012}*** & \num{0.019}*** & \num{0.013}*** \\
 & (\num{0.004}) & (\num{0.004}) & (\num{0.004}) & (\num{0.003}) & (\num{0.003}) \\
t = 1                & \num{0.036}*** & \num{0.029}*** & \num{0.029}*** & \num{0.038}*** & \num{0.031}*** \\
 & (\num{0.004}) & (\num{0.004}) & (\num{0.004}) & (\num{0.003}) & (\num{0.004}) \\
\midrule
\textbf{Mean conversion rate} & \multicolumn{3}{c|}{0.043} & \multicolumn{2}{c}{0.041} \\
Proportional change wrt mean  & 0.30 & 0.35 & 0.35 & 0.27 & 0.32 \\
\midrule
Num. Obs.  & \num{76375} & \num{76375} & \num{76375} & \num{113576} & \num{113576} \\
Std. Errors& IID & IID & by: Wrkpl & IID & by: Wrkpl \\
Covariates &  & \checkmark & \checkmark &  & \checkmark \\
\bottomrule
\end{tabular}
}
\footnotesize
\vspace{0.1cm}\\
Standard errors are in parentheses.\\
Significance level: $^{*} p < 0.1$, $^{**} p < 0.05$, $^{***} p < 0.01$
\end{table}

\begin{figure}[ht]
    \centering
    \includegraphics[width=0.8\linewidth]{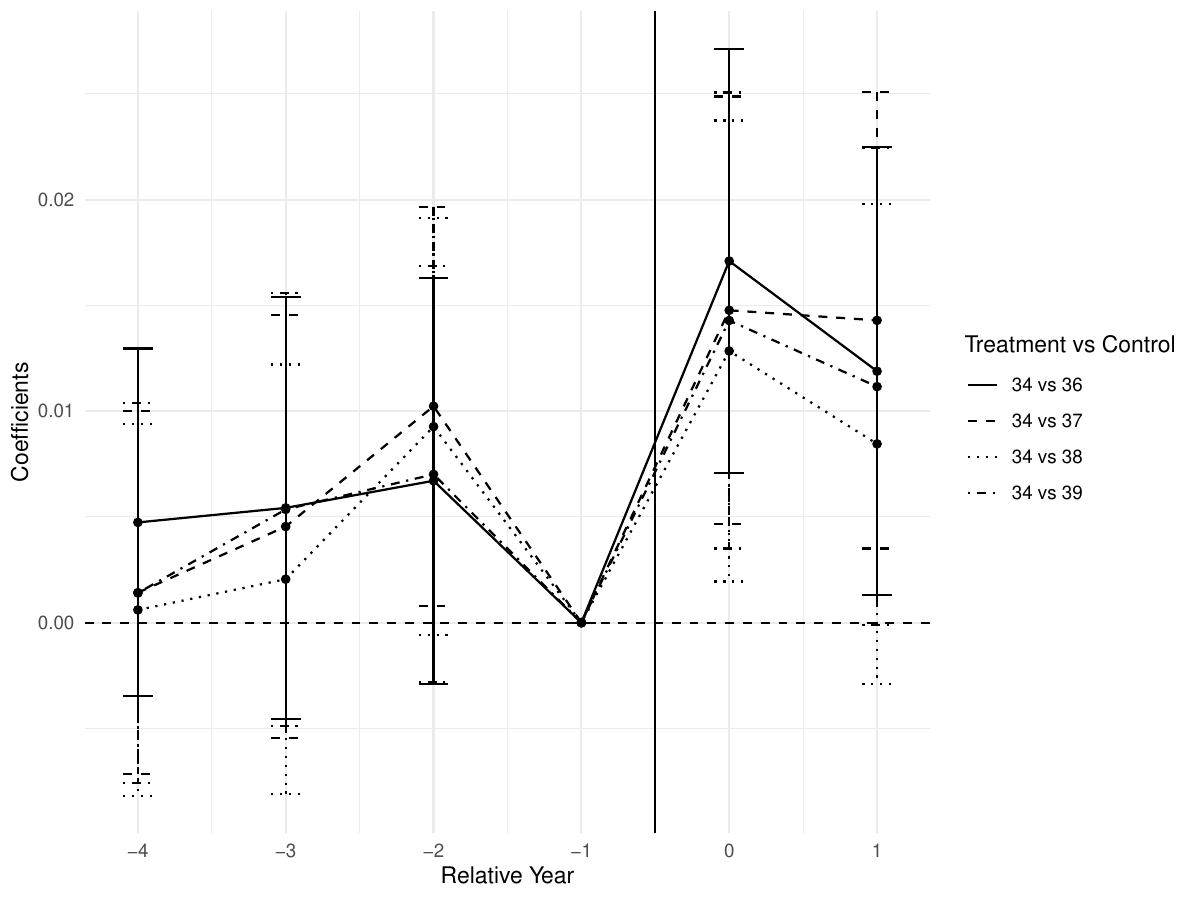}\caption{\textsc{DiD spillover check} \\ The plot shows the coefficients and 95\% confidence intervals for the DiD regressions, where the control group consists of individuals aged 36 to 39. }
    \label{fig:figure4}
\end{figure}

\paragraph{Spillover effects} We now turn to studying the substitution between eligible and non-eligible workers. We begin by studying whether our results were sensitive to the choice of the control group. For this purpose, we rely on the standard binary DiD specification, which ensures greater robustness while also introducing alternative control groups encompassing individuals aged 37 to 39. This approach allows us to test whether non-eligible workers just above the threshold, being the most similar to the eligible group, experienced a differential impact compared with older cohorts. We find that the estimated effects remain very similar, as can be easily noted from the estimates plotted in \hyperref[fig:figure4]{Figure~\ref{fig:figure4}}.  Parallel trends also generally hold among all cohorts, with the only exception of a small difference in hiring rates between the 34 and 37-year-olds in 2016. These results suggest the absence of spillover effects, as 36-year-olds are not affected differently from their slightly older peers. The estimated coefficients are shown in Appendix \ref{Appendix:did}, in the first three columns of \hyperref[tab:control_spillover]{Table~\ref{tab:control_spillover}}.

As a final check, we look at the within-workplace workforce composition of workers of similar age. Specifically, we look at the share of eligible co-workers among the 34 and 36 age cohorts in the same workplace the year before the observation. 
We then interact this indicator with our treatment variable, following the same approach as in the baseline model, to assess whether conversion rates differed systematically in firms that hired a higher proportion of 34-year-olds compared to those employing slightly older cohorts (i.e., the 36-year-olds). The estimates reported in the last column of \hyperref[tab:control_spillover]{Table~\ref{tab:control_spillover}} show no statistically significant effect, suggesting that the observed increase in conversions was not linked with opportunistic adjustments in hiring practices but rather reflected genuine take-up of the incentive among firms already employing eligible workers.

\paragraph{Heterogeneity} 
Detailed heterogeneity results are reported in Appendix~\ref{Appendix:heterogeneity} and summarized in \hyperref[fig:griglia]{Figure~\ref{fig:griglia}}. As expected, fixed-term contracts in the industrial and service sectors show a higher propensity to be converted into permanent positions compared to those in agriculture (Panel a, \hyperref[fig:griglia]{Figure~\ref{fig:griglia}}). This finding is consistent with the notion that industrial and service firms, which are typically characterized by more stable demand for labor, were more likely or better positioned to take advantage of the subsidy by stabilizing their under-35 employees. 
Regarding skill intensity, Panel b of the same figure shows that the positive impact on conversions is mainly driven by medium-skill occupations.

To explore the role of firm size, we approximate this dimension by examining the distribution of hirings across years and sectors, using the 75th percentile as a threshold to distinguish between smaller and larger firms. Under this definition, the positive effect of the policy appears to be concentrated primarily among smaller firms, whereas virtually no significant effect is detected for larger firms (Panel c, \hyperref[fig:griglia]{Figure~\ref{fig:griglia}}). A plausible explanation is that larger firms may have perceived the subsidy as insufficiently generous or economically meaningful to alter their hiring behavior.

Contract history also proves to be an important factor. Contracts whose potential duration exceeds the median and are therefore eligible for conversion over a longer period show a significantly higher probability of being transformed into permanent ones (Panel d). 
This result is consistent with a longer window for firms to screen productivity prior to conversion and with higher effective treatment intensity (i.e., more days on fixed-term contracts, resulting in greater exposure to the incentive).

Finally, when considering gender, the results indicate that the increase in conversion rates is almost entirely driven by male workers (Panel e).

\FloatBarrier

\subsubsection{Regression Discontinuity and RD-DiD estimation}
To complement the DiD analysis, we also implement a Regression Discontinuity Design that exploits the sharp change in eligibility at the 35th birthday, providing a local causal estimate of the policy effect. This time, we consider only the people turning 35 in 2018 in a symmetric 3-month window around the threshold, in the days in which they were in a temporary contract. Thus, the running variable is defined as the distance in days from the individual’s birthday. 

\begin{figure}[ht]
        \centering
        \includegraphics[width=0.8\linewidth]{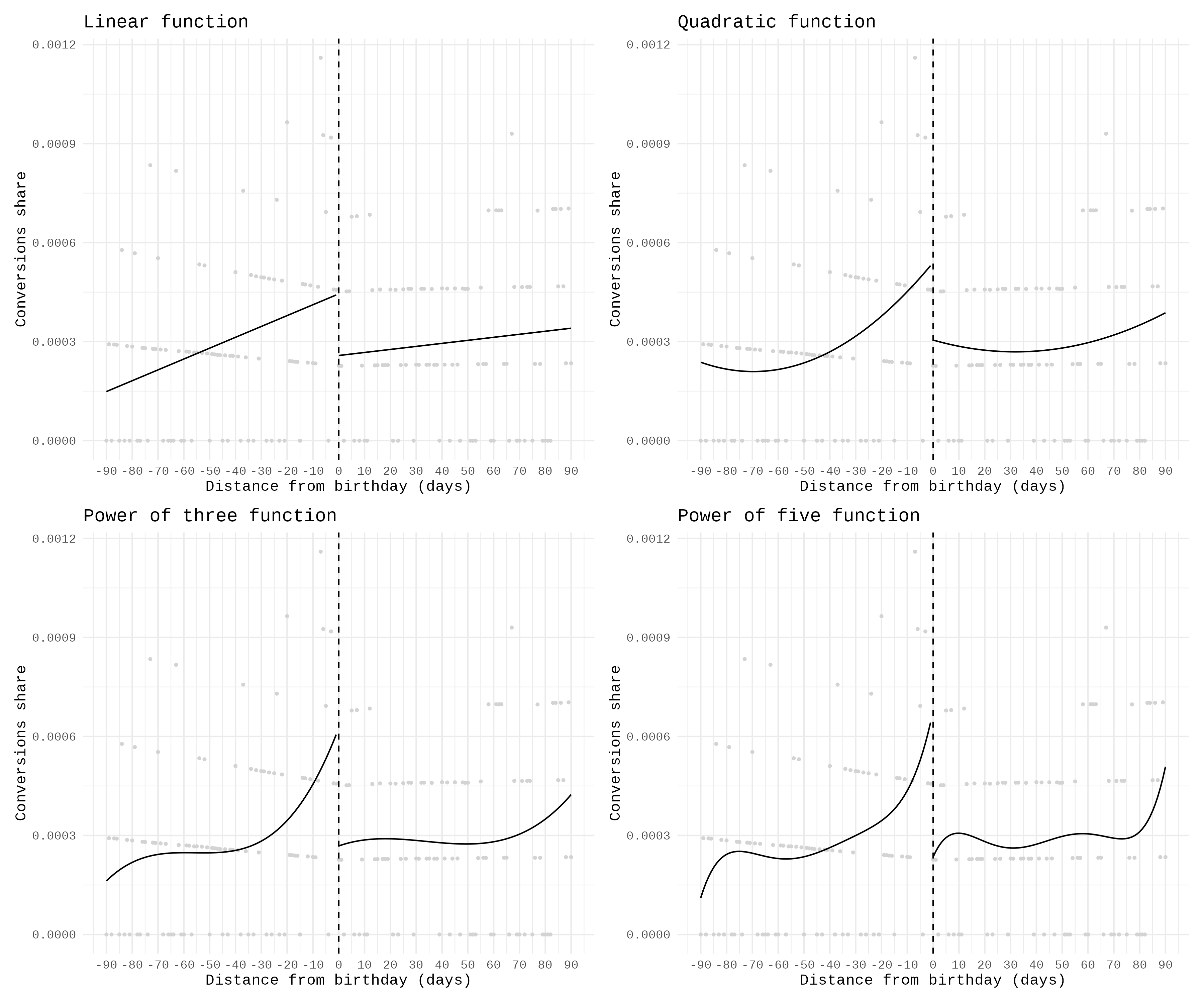}
        \caption{\textsc{Approximation of the functional form of the running variable} \\   The plot presents descriptive outcomes from fitting the running variable with linear, quadratic, cubic, and fifth-degree polynomial functions.}
        \label{fig:figure6}
    \end{figure}

As discussed in the econometric section, we first choose to approximate the running variable function using both linear and quadratic specifications. As shown in \hyperref[fig:fig2]{Figure~\ref*{fig:figure6}}, the functional form appears to stabilize beyond the second degree, suggesting that higher-order polynomials may not add meaningful explanatory power and lead to overfitting.

In our setting, the running variable is the worker’s age in days around the 35th birthday. Since this varies continuously within individuals and cannot be manipulated by workers or firms, concerns about strategic sorting at the threshold are inherently ruled out. A downward slope nevertheless appears on the left-hand side of the density plots. This pattern arises mechanically from truncation: birthdays earlier in the year are more likely to generate observations to the left of the threshold that fall into the previous calendar year (2018), and these are dropped. As a result, the number of individuals decreases with distance from the cutoff on the left-hand side. Including such observations would have introduced bias, as some individuals would have been mistakenly classified as eligible when they were not.
The McCrary density test, presented in Section~\ref{s:robustness}, confirms continuity of the running variable at the threshold, supporting the validity of our design. 

\ref{tab:table_4} presents the results of a simplified regression discontinuity design on the whole 90-day window. Columns 1 and 2 estimate the model using a linear specification of the running variable, while columns 3 and 4 adopt a quadratic one. Furthermore, in the second and fourth columns, we add the same covariates used for the DiD model. Across all specifications, the estimated effects are positive and statistically significant, and their magnitude remains stable regardless of covariate inclusion.
Moreover, the quadratic specification yields results that are highly consistent with the linear one, albeit with slightly reduced statistical significance, likely due to its greater flexibility. Although the estimated coefficients are numerically small, they represent a substantial effect: when expressed relative to the average daily probability of conversion, they correspond to an increase ranging from 67\% to 81\% around the threshold. As expected, the RD estimate of the policy's effect is higher because it captures the sharp response at the cutoff birthday, whereas the DiD averages outcomes annually across workers who differ in exposure intensity, as treatment applies only on days in fixed-term jobs.
\begin{table}[ht]
\centering
\title{\textbf{Table 2: Regression Discontinuity Design estimation}}
\caption{}
\label{tab:table_4}
\resizebox{1\textwidth}{!}{
\begin{tabular}{lcc|cc}
\toprule
\midrule
 & \multicolumn{2}{c}{Linear Approximation} 
 & \multicolumn{2}{c}{Quadratic Approximation}\\
\cmidrule(lr){2-3} \cmidrule(lr){4-5}
 & Model 1 & Model 2 &  Model 3 & Model 4 \\
\midrule

Eligible         & 0.0002$^{**}$  & 0.00021$^{**}$  & 0.00025$^{*}$  & 0.00024$^{*}$  \\
                & (0.00008)       & (0.00009)       & (0.00013)      & (0.00013)      \\
\midrule
\makecell{Mean conversion rate: 0.0003} &&&&\\
Proportional change wrt mean & 0.67 & 0.69 & 0.83 & 0.80\\
\midrule
Bandwidth: 90 days &&&&\\
Num. Obs.       & 736,491          & 736,491          & 736,491         & 736,491         \\
Covariates && \checkmark& &\checkmark\\
\bottomrule
\end{tabular}
}
\footnotesize 
\vspace{0.1cm}\\
All the standard errors are clustered at workplace and individual level. \\
Significance level: $^{*} p < 0.1$, $^{**} p < 0.05$, $^{***} p < 0.01$
\end{table}

However, following \cite{Cattaneo2019}, nowadays it is widely recognized that global polynomial approaches do not deliver point estimators and inference procedures with good properties for the treatment effect, because they tend to deliver a good approximation overall, but a poor one at boundary points. The authors recommend implementing local polynomial methods, focusing only on the region near the cutoff, discarding observations sufficiently far away, and employing a low-order polynomial approximation (usually linear or quadratic). This approach is less sensitive to boundary and overfitting problems.

Following this framework, we implement a linear regression using the Optimal Bandwidth Selection (OPS) procedure \citep{Calonico2019} with a triangular kernel.\footnote{These estimates are obtained using the package \textit{rdrobust} developed by the same authors \citep{RJ-2015-004}.} The results are reported in \ref{tab:table_5}. The OPS procedure selected a window length of 36 days around the cutoff, and the associated coefficient is substantially higher with respect to the previous global estimation. 
The resulting coefficient is statistically different from 0 at a 90\% confidence level and corresponds to an increase of 92\% with respect to the conversion probability in that region. Furthermore, we also report the results of the regression estimated with the inclusion of the covariates, with both a symmetrical (second row) and asymmetrical (third row) bandwidth. The estimated coefficients are still almost identical to the simplest specification, both in magnitude and significance level.


\begin{table}[ht]
\centering
\title{\textbf{TABLE 3: Regression discontinuity with Optimal Bandwidth and Triangular Kernel approximation}}
\caption{}
\label{tab:table_5}
\resizebox{\textwidth}{!}{
\begin{tabular}{rlrrrrr}
\toprule
\midrule
 Bandwidth & Covariates &  Coefficient &  Robust SE & Pvalue & \makecell{Proportional effect\\wrt mean} \\
\midrule
35.6 &  & 0.00031$^{*}$ & 2e-04 &  0.080 & 0.92 \\
35.6 &   \multicolumn{1}{c}{\checkmark}  & 0.00031$^{*}$ & 2e-04 &  0.078 & 0.92 \\
35.2 (L) 33.5(R) & \multicolumn{1}{c}{\checkmark}  & 0.00034$^{*}$ & 2e-04 &  0.082 & 1.01 \\
\bottomrule
\end{tabular}
}
\footnotesize
\vspace{0.1cm}\\
All the standard errors are clustered at workplace and individual level\\
Significance level: $^{*} p < 0.1$, $^{**} p < 0.05$, $^{***} p < 0.01$
\end{table}

Building on this, we further estimate an RD-DiD specification \citep{Grembi2016}, in which we exploit the same discontinuity but introduce additional cohorts of individuals who turned 35 in pre-reform years. This allows us to test whether any observed discontinuity at the threshold is genuinely attributable to the 2018 policy, rather than reflecting pre-existing structural breaks around age 35. In practice, we estimate the RD-DiD within a symmetric 35-day bandwidth around the cutoff, consistent with the optimal bandwidth selection obtained in the standard RD analysis, and introduce interaction terms with the year relative to the reform.

\ref{tab:table_6} reports the coefficients from both a baseline model and a specification including covariates, with standard errors clustered at the individual and workplace level.
The results show a clear and robust discontinuity in the year of policy implementation (t = 0). The coefficient is positive, statistically significant at the 5\% level, and implies a proportional increase of around 1.5 times the mean conversion rate, highlighting that the reform substantially increased the probability of conversion precisely at the eligibility margin. Crucially, the estimates for the pre-reform cohorts (t = -4, -3, -2) are small in magnitude and statistically indistinguishable from zero, providing compelling evidence that no discontinuity existed in earlier years. This supports the validity of the RD design and reinforces the interpretation that the effect observed in 2018 is a genuine policy impact, rather than a spurious age-related discontinuity. Furthermore, as in the DiD estimation, the coefficient for t = 1 (the year after the reform) remains positive and marginally significant, again suggesting anticipation of the extension of the threshold that occurred at the end of this year. 

\begin{table}[h!]
\centering
\title{\textbf{table 4: Difference in Discontinuity estimation}}
\caption{}
\label{tab:table_6}
\begin{tabular}{llcc}
\toprule
\midrule
 & Model 1 &  Model 2 \\
\midrule
Eligible $\times$  (t = 0) & 0.00045$^{**}$ & 0.00045$^{**}$ \\
            & (0.00020)       & (0.00020) \\
Eligible $\times$  (t = -4) & 0.00008 & 0.00007 \\
            & (0.00016)       & (0.00016) \\
Eligible $\times$  (t = -3) & 0.00021 & 0.00021 \\
            & (0.00018)       & (0.00018) \\
Eligible $\times$  (t = -2) & 0.00007 & 0.00007 \\
            & (0.00013)       & (0.00013) \\
Eligible $\times$  (t = 1) & 0.00035$^{*}$ & 0.00035$^{*}$ \\
            & (0.00019)       & (0.00019) \\
Eligible     & -0.00013        & -0.00012 \\
            & (0.00010)       & (0.00010) \\
t = -4 & 0.00000      & -0.00002 \\
            & (0.00012)       & (0.00012) \\
t = -3 & 0.00006      & 0.00005 \\
            & (0.00012)       & (0.00012) \\
t = -2 & -0.00012     & -0.00012 \\
            & (0.00010)       & (0.00010) \\
t = 0 & 0.00009       & 0.00008 \\
            & (0.00012)       & (0.00012) \\
t = 1 & 0.00014       & 0.00014 \\
            & (0.00012)       & (0.00012) \\
            \midrule
Num.Obs.   & \num{1385129} & \num{1385129} \\
Covariates && \checkmark\\
\bottomrule
\end{tabular}
\footnotesize
\vspace{0.1cm}\\
All the standard errors are clustered at workplace and individual level\\
Significance level: $^{*} p < 0.1$, $^{**} p < 0.05$, $^{***} p < 0.01$
\end{table}

\paragraph{Spillover effects}  We look again at spillover effects, restricting the sample to firms employing at least two workers turning 35 years old in 2018 (for the RD specification) or in each observation year (for the RD-DiD). For each individual, we construct the spillover variable $S$ as detailed in Section \ref{s:method} and interact it with the treatment and running variable. Our results are reported in Appendix \ref{Appendix:RDspillover}, \hyperref[tab:rdd_rdddid]{Table~\ref{tab:rdd_rdddid}}.

The coefficients of primary interest are $S \times Eligible$ (in the RD) and $S \times Eligible \times (t=0)$ (in the RD-DiD), and are consistently negative. However, these estimates are statistically insignificant across all specifications, suggesting that the presence of a larger pool of ineligible coworkers does not systematically increase the probability of conversion. 

\FloatBarrier


\subsection{Post-policy effects}
In our medium-term analysis, we investigate whether the short-run increase in permanent contract conversions translated into more durable improvements in employment trajectories. \ref{tab:table_7} presents the results of the standard DiD estimation, where individuals just below and above the eligibility cutoff (aged 34 vs. 36) are compared one year after treatment. Contrary to the short-run findings, we do not detect any statistically significant impact of the policy on subsequent career outcomes. In particular, the probability of eligible individuals being employed under an open-ended contract does not appear to increase in the year following conversion, despite the sharp rise in conversions observed in the immediate aftermath of the reform. This pattern is consistent with the idea that firms may have simply anticipated conversions that would have occurred later on, taking advantage of the temporary incentives.

Turning to longer-term horizons, \ref{tab:table_8} examines outcomes up to four years after the reform, while analogous estimates for two and three years are provided in the Appendix \ref{Appendix:medium-run}, Table \ref{tab:med_2} e \ref{tab:med_3}. Once again, we find no significant policy effects on medium-run employment stability. This suggests that the initial increase in conversions largely failed to consolidate into more persistent gains in terms of open-ended employment relationships. 

Nevertheless, one reassuring result emerges from the analysis. Even four years after the reform, we do not observe a reversal effect, i.e., a systematically lower probability of maintaining an open-ended contract among the treated group. Such an outcome would have raised concerns that firms converted workers solely to capture the subsidy, only to terminate the relationship at the earliest opportunity. The absence of such a pattern is notable, especially in the context of Italy’s \textit{contratto a tutele crescenti}, introduced in 2015, which initially reduced employment protection for newly hired permanent workers and could have facilitated early dismissals. Our findings therefore suggest that, although the policy did not generate durable improvements in employment trajectories, it also did not produce perverse incentives leading to systematically shorter employment spells among treated workers.

\begin{table}[!ht]
\centering
\title{\textbf{TABLE 5: DiD medium-run outcomes one year after}}
\begin{tabular}[t]{llllll}
\toprule
\midrule
 & \makecell{Same \\ contract} & \makecell{Permanent \\ contract} & \makecell{Same \\ firm} & \makecell{Same \\ sector} & \makecell{Days \\ worked}\\
\midrule
Eligible $\times$  (t=0)  & \makecell{-0.005 \\ (0.014)} & \makecell{0.018 \\ (0.014)}  & \makecell{0.015 \\ (0.014)}  & \makecell{0.016 \\ (0.014)}  & \makecell{0.012 \\ (0.012)}\\
Eligible $\times$  (t=-4) & \makecell{-0.014 \\ (0.013)} & \makecell{-0.006 \\ (0.015)} & \makecell{-0.007 \\ (0.014)} & \makecell{-0.007 \\ (0.014)} & \makecell{-0.002 \\ (0.013)}\\
Eligible $\times$  (t=-3) & \makecell{-0.019 \\ (0.014)} & \makecell{-0.007 \\ (0.015)} & \makecell{-0.002 \\ (0.015)} & \makecell{-0.006 \\ (0.015)} & \makecell{-0.001 \\ (0.013)}\\
Eligible $\times$  (t=-2) & \makecell{-0.017 \\ (0.015)} & \makecell{-0.021 \\ (0.014)} & \makecell{-0.019 \\ (0.014)} & \makecell{-0.022 \\ (0.014)} & \makecell{-0.015 \\ (0.013)}\\
t=-4 & \makecell{-0.027* \\ (0.012)} & \makecell{-0.001 \\ (0.011)} & \makecell{0.002 \\ (0.011)} & \makecell{0.001 \\ (0.011)} & \makecell{0.014 \\ (0.009)}\\
t=-3 & \makecell{-0.040*** \\ (0.012)} & \makecell{0.020* \\ (0.011)} & \makecell{0.021* \\ (0.011)} & \makecell{0.017 \\ (0.011)} & \makecell{0.014 \\ (0.010)}\\
t=-2 & \makecell{-0.041*** \\ (0.012)} & \makecell{-0.032*** \\ (0.011)} & \makecell{-0.031*** \\ (0.011)} & \makecell{-0.038*** \\ (0.011)} & \makecell{-0.033*** \\ (0.009)}\\
t=0  & \makecell{0.012 \\ (0.010)}     & \makecell{0.012 \\ (0.010)} & \makecell{0.016 \\ (0.010)} & \makecell{0.019* \\ (0.010)} & \makecell{0.019** \\ (0.009)}\\
Eligible     & \makecell{0.018* \\ (0.010)}    & \makecell{0.011 \\ (0.010)} & \makecell{0.010 \\ (0.010)} & \makecell{0.013 \\ (0.010)} & \makecell{0.007 \\ (0.009)}\\
\midrule
Num. Obs. & 63646 & 35683& 35683&35683&35683\\
\bottomrule
\end{tabular}
\footnotesize
\vspace{0.1cm}\\
All the standard errors are clustered at workplace level\\Significance level: $^{*} p < 0.1$, $^{**} p < 0.05$, $^{***} p < 0.01$
\caption{}
\label{tab:table_7}
\end{table}

\begin{table}[!ht]
\centering
\title{\textbf{TABLE 6: DiD medium-run outcomes four years after}}
\begin{tabular}[t]{llllll}
\toprule
\midrule
 & \makecell{Same \\ contract} & \makecell{Permanent \\ contract} & \makecell{Same \\ firm} & \makecell{Same \\ sector} & \makecell{Days \\ worked}\\
 \midrule
 Eligible $\times$ (t=0)  & \makecell{0.003 \\ (0.007)} & \makecell{0.006 \\ (0.015)} & \makecell{0.006 \\ (0.015)} & \makecell{0.002 \\ (0.014)} & \makecell{0.004 \\ (0.011)}\\
 Eligible $\times$  (t=-4)& \makecell{0.001 \\ (0.006)} & \makecell{0.001 \\ (0.015)} & \makecell{0.001 \\ (0.015)} & \makecell{0.001 \\ (0.014)} & \makecell{0.003 \\ (0.011)}\\
Eligible $\times$  (t=-3) & \makecell{0.001 \\ (0.006)} & \makecell{0.019 \\ (0.016)} & \makecell{0.023 \\ (0.016)} & \makecell{0.017 \\ (0.015)} & \makecell{0.012 \\ (0.012)}\\
Eligible $\times$  (t=-2) & \makecell{-0.002 \\ (0.006)} & \makecell{-0.017 \\ (0.016)} & \makecell{-0.015 \\ (0.015)} & \makecell{-0.023 \\ (0.015)} & \makecell{-0.012 \\ (0.012)}\\
t=-4 & \makecell{-0.034*** \\ (0.004)} & \makecell{-0.021* \\ (0.012)} & \makecell{-0.024** \\ (0.011)} & \makecell{-0.040*** \\ (0.011)} & \makecell{-0.028*** \\ (0.009)}\\
t=-3 & \makecell{-0.025*** \\ (0.005)} & \makecell{0.000 \\ (0.012)}  & \makecell{-0.008 \\ (0.011)} & \makecell{-0.024** \\ (0.011)} & \makecell{-0.020** \\ (0.009)}\\
t=-2 & \makecell{-0.022*** \\ (0.005)} & \makecell{-0.011 \\ (0.012)} & \makecell{-0.012 \\ (0.011)} & \makecell{-0.024** \\ (0.011)} & \makecell{-0.012 \\ (0.008)}\\
t=0  & \makecell{0.007 \\ (0.005)}     & \makecell{0.005 \\ (0.011)} & \makecell{0.003 \\ (0.011)} & \makecell{0.001 \\ (0.010)} & \makecell{-0.009 \\ (0.008)}\\
Eligible     & \makecell{0.006 \\ (0.005)}     & \makecell{0.013 \\ (0.011)} & \makecell{0.013 \\ (0.011)} & \makecell{0.019* \\ (0.010)} & \makecell{0.014* \\ (0.008)}\\
\midrule
Num. Obs. & 63646 & 35683& 35683&35683&35683\\
\bottomrule
\end{tabular}
\footnotesize
\vspace{0.1cm}\\
All the standard errors are clustered at workplace level\\Significance level: $^{*} p < 0.1$, $^{**} p < 0.05$, $^{***} p < 0.01$
\caption{}
\label{tab:table_8}
\end{table}

\paragraph{Heterogeneity} We repeat our heterogeneity analysis with respect to the medium-term outcomes to assess whether policy effects re-emerge when mediated by other employers and employee characteristics. However, across all outcomes and time horizons, no clear pattern appers, as the vast majority of estimates show no significant differences. In \hyperref[fig:het2]{Figure~\ref{fig:het2}} and \hyperref[fig:het3]{Figure~\ref{fig:het3}} (Appendix~\ref{Appendix:heterogeneity}), we report the results of the heterogeneity analysis for one year and four years forward, respectively.

\FloatBarrier

\subsection{Robustness checks}
\label{s:robustness}
In this section, we report a set of robustness checks.

\begin{figure}[!ht]
    \centering
    \includegraphics[width=0.8\linewidth]{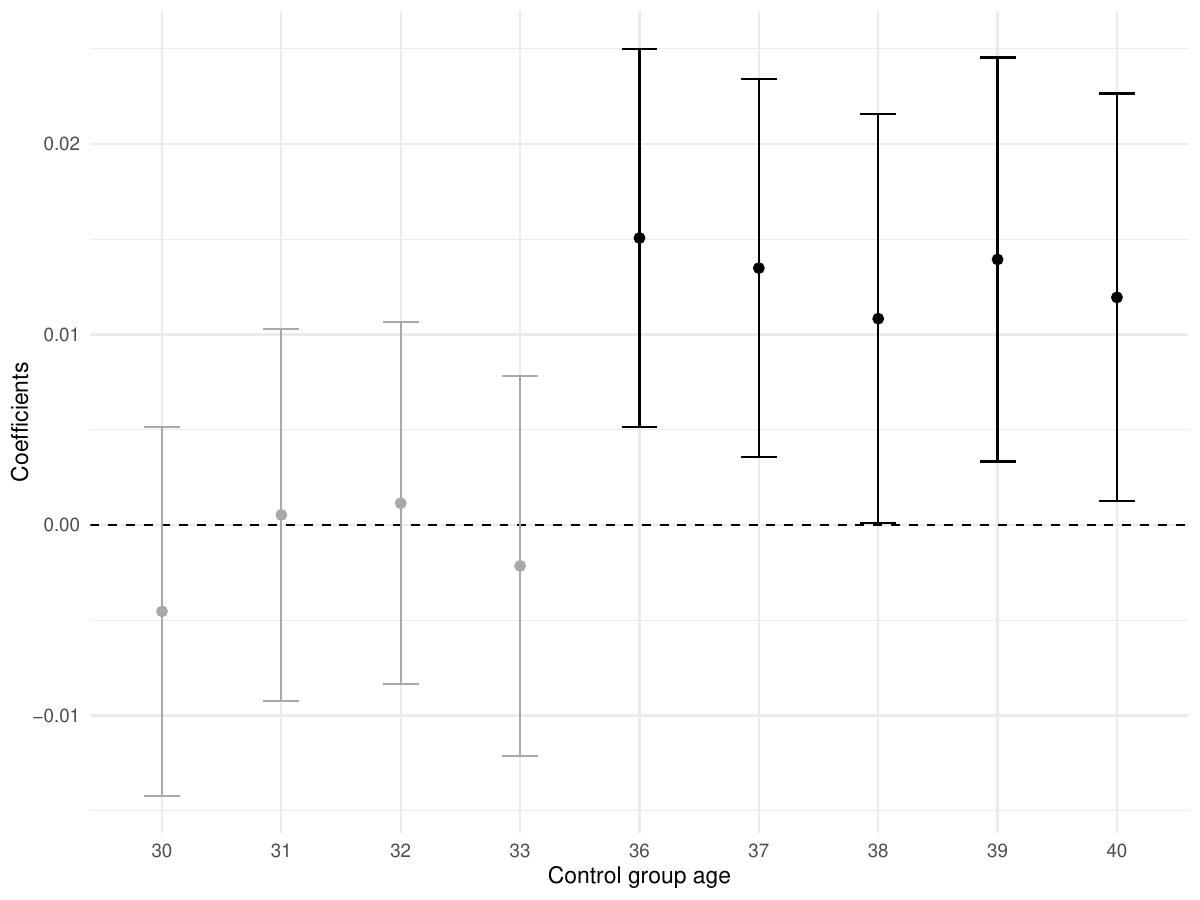}
    \caption{\textsc{Placebo test with fake threshold - DiD} \\  The plot reports the coefficients and the 95\% confidence intervals of the DiD regressions estimated using the 34 years-old cohorts as the treatment group but changing the control group from people age 30 to 40 (in the $\times$ axis).}
    \label{fig:falsification_1}
\end{figure}

We begin with the DiD estimations. As a falsification exercise, we extend our approach to spillover effects using the 34 years-old cohorts as the treatment group but switching the control group with individuals from cohorts varying, in each specification, from age 30 to 40. This is a standard test to ensure that no effect is found when we compare the 34-year-olds with younger eligibile cohorts. The results are reported in \hyperref[fig:fig1]{Figure~\ref*{fig:falsification_1}}. The only significant policy effects arise when comparing 34-year-olds with untreated workers aged 36 years or older. Furthermore, as already highlighted in the spillover section, the effect is also consistently similar in size across all specifications. 

As a further check, we also decided to examine whether any effect emerges around the 30-year-old threshold, which was not binding until 2019. If the identification strategy based on the 35-year-old limit is correct, no effect should be found for this group in 2018, while it can arise in the following year given that this threshold remained in place until December of that year. However, as shown in \hyperref[fig:fig1]{Figure~\ref*{fig:contvsstand}}, we find no statistically significant results for conversions in this age group in either 2018 or 2019. For the latter, the coverage for individuals over 30 was announced only with the Financial Law, which was issued and published at the end of the year. The zero coefficient for this year can mean that firms took advantage of the larger eligibility criteria of the previous year, converting more older people. Alternatively, employers might have anticipated the retroactive extension of coverage for older individuals, which would also explain why we found an increase around the 35-year-old threshold. Lastly, the policy was ineffective for this age group, who were also covered by other policies (like the YG hiring incentive). It is important to note that the failing of the parallel trend for the relative year -3 and -4 is probably due to the effect of the untargeted incentives implemented under the Jobs Act.

\begin{figure}[!ht]
    \centering
    \includegraphics[width=0.8\linewidth]{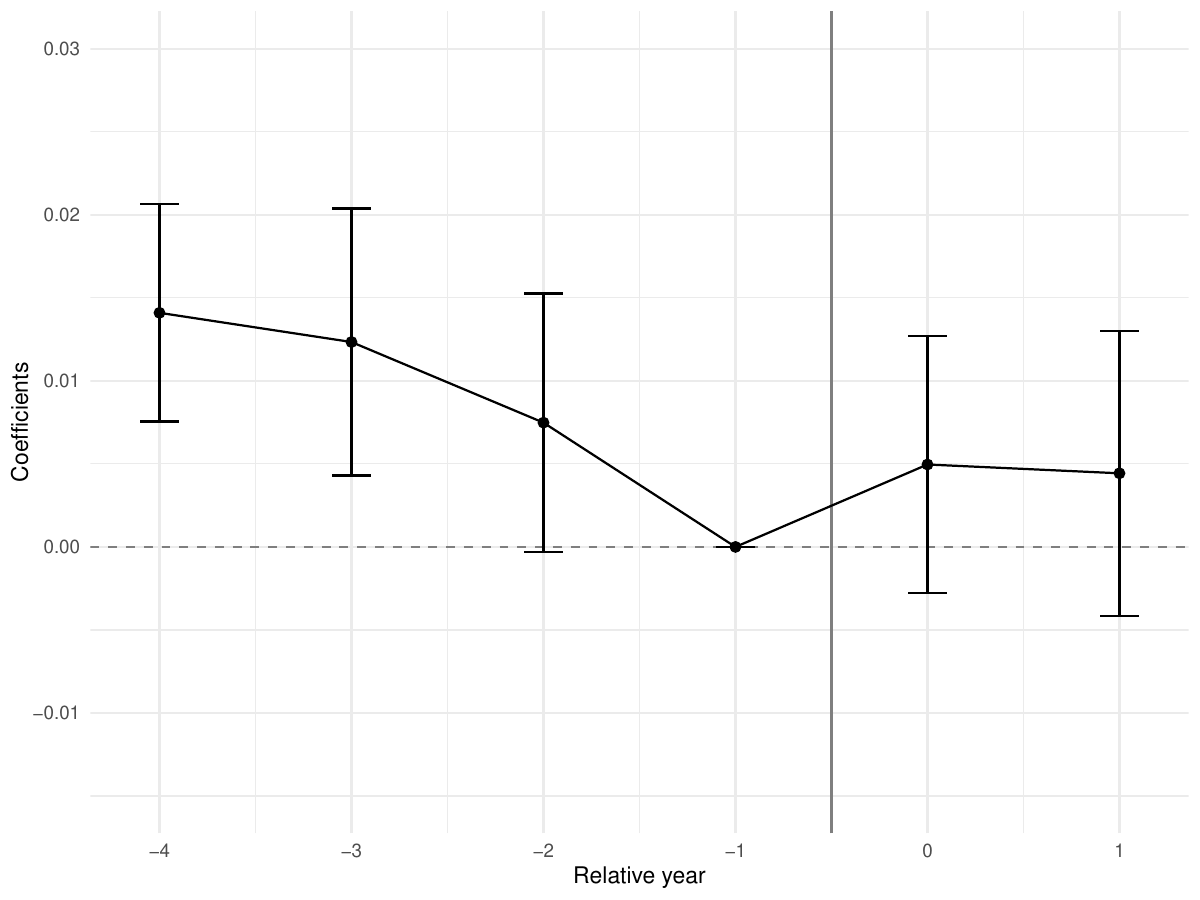}
    \caption{\textsc{DiD regression with fake cutoff at 30 years old} \\  In this plot are represented the coefficients of the DiD using people tuning 29 as treated and 30 as controls. The error bars represent the 95\% confidence intervals.}
    \label{fig:falsification}
\end{figure}

\begin{figure}[!ht]
    \centering
    \includegraphics[width=0.8\linewidth]{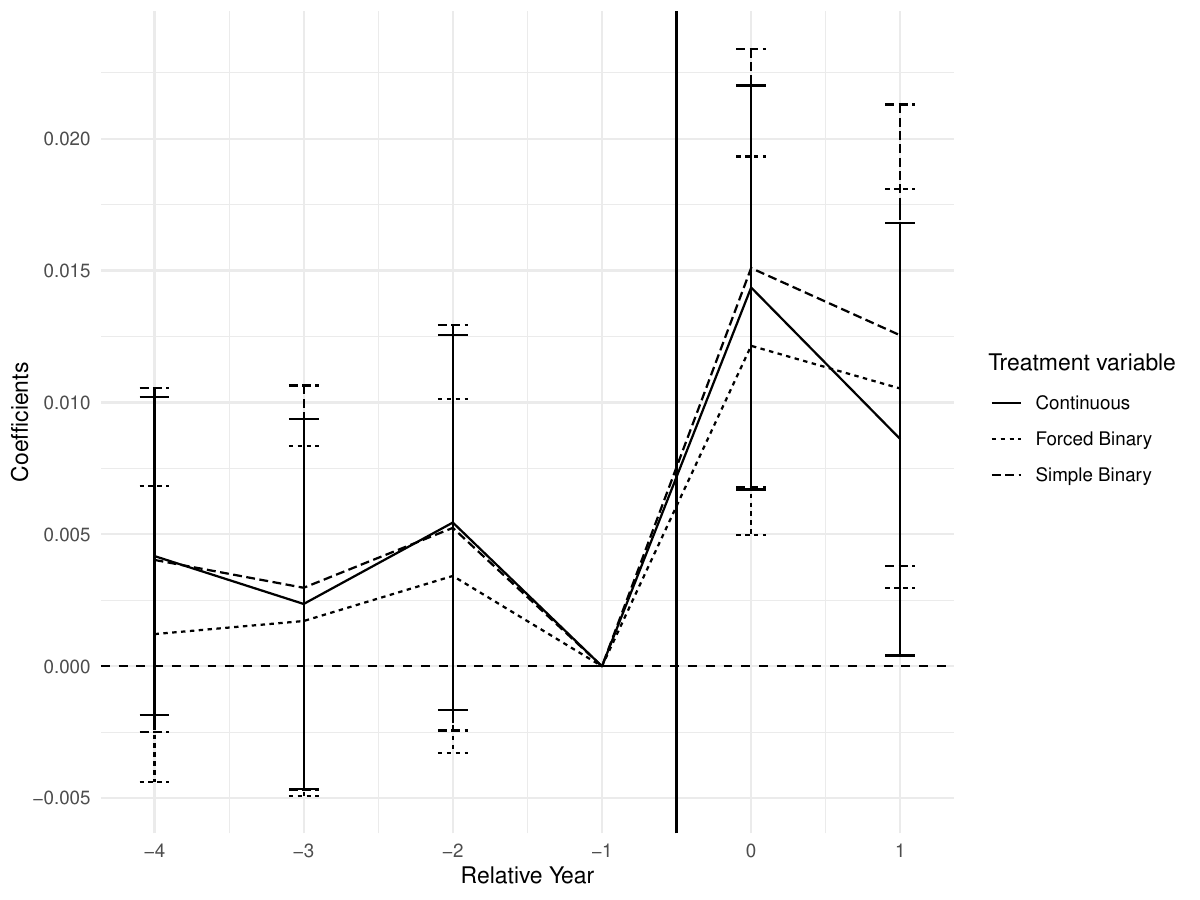}
    \caption{\textsc{DiD estimates comparison} \\In this plot are represented the coefficients of the DiD, including only 34 and 36-year-old individuals (Simple Binary), and the ones corresponding to the whole sample and the treatment considered as continuous (Continuous), and lastly, the whole sample and the treatment forced to be binary (Forced binary). The error bars represent the 95\% confidence intervals.}
    \label{fig:contvsstand}
\end{figure}

\FloatBarrier

\begin{figure}[!ht]
    \centering
    \includegraphics[width=0.8\linewidth]{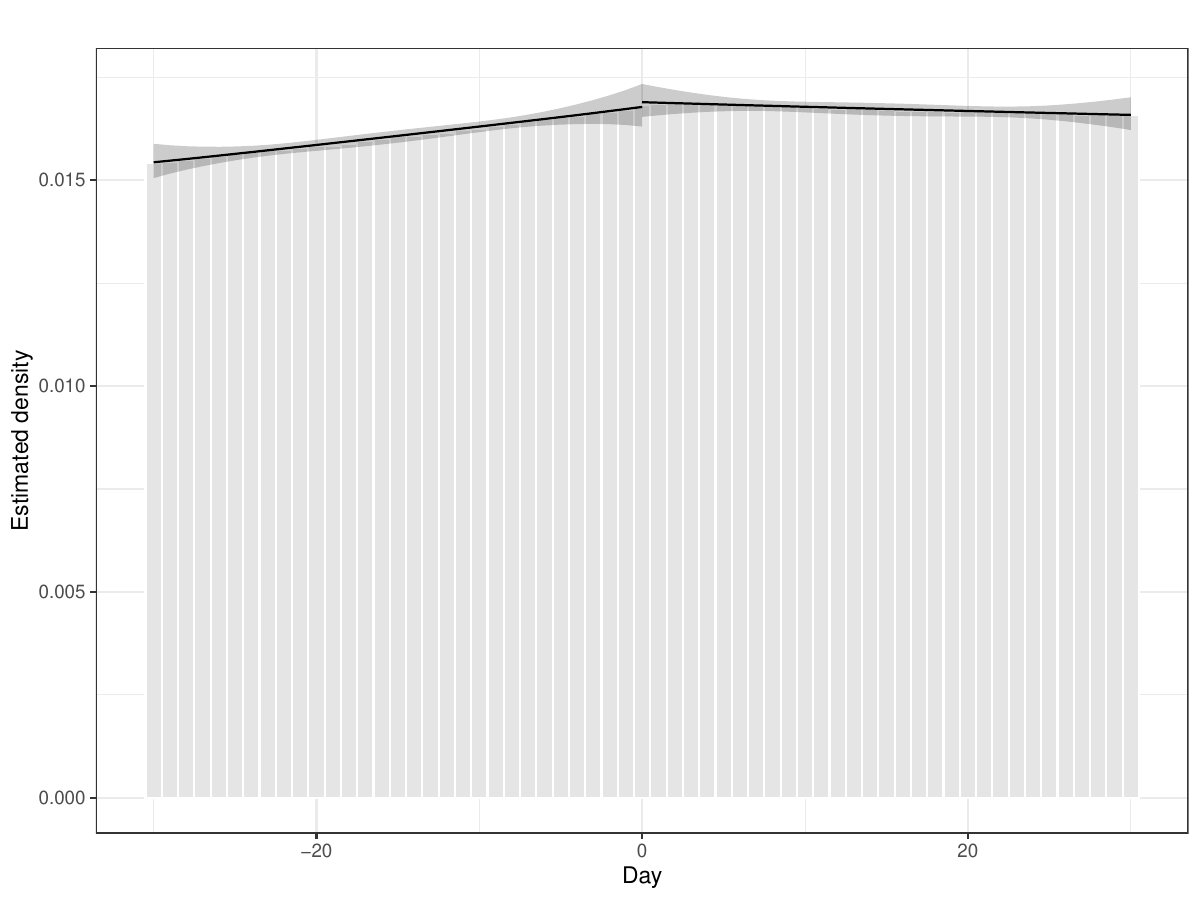}
    \caption{\textsc{Density test for the RD} \\ The plot shows the distribution of the observations in the RD sample, with 95\% confidence interval.}
    \label{fig:density}
\end{figure}

Moving to the regression discontinuity design, to assess the robustness, we conduct a density check of the running variable around the cutoff. Specifically, we employ the \textit{rddensity} function in R, which allows us to visually inspect the distribution of the running variable and identify any potential discontinuities near the threshold. This approach is complemented by the McCrary test, which provides a formal statistical test for any manipulation of the running variable. The absence of significant jumps in the density at the cutoff indicates that observations are continuously distributed around the threshold, thereby supporting the internal validity of our RDD. The density check is presented in \hyperref[fig:density]{Figure~\ref*{fig:density}}, providing an intuitive visual assessment and showing that the data are balanced just above and below the cutoff. 
Similar figures are also emerge for the entirety of the RD-DiD estimation window in \hyperref[fig:figDENSITY]{Figure~\ref*{fig:figDENSITY}}, Appendix  \ref{Appendix:RDspillover}.

\begin{figure}[!ht]
    \centering
    \includegraphics[width=0.8\linewidth]{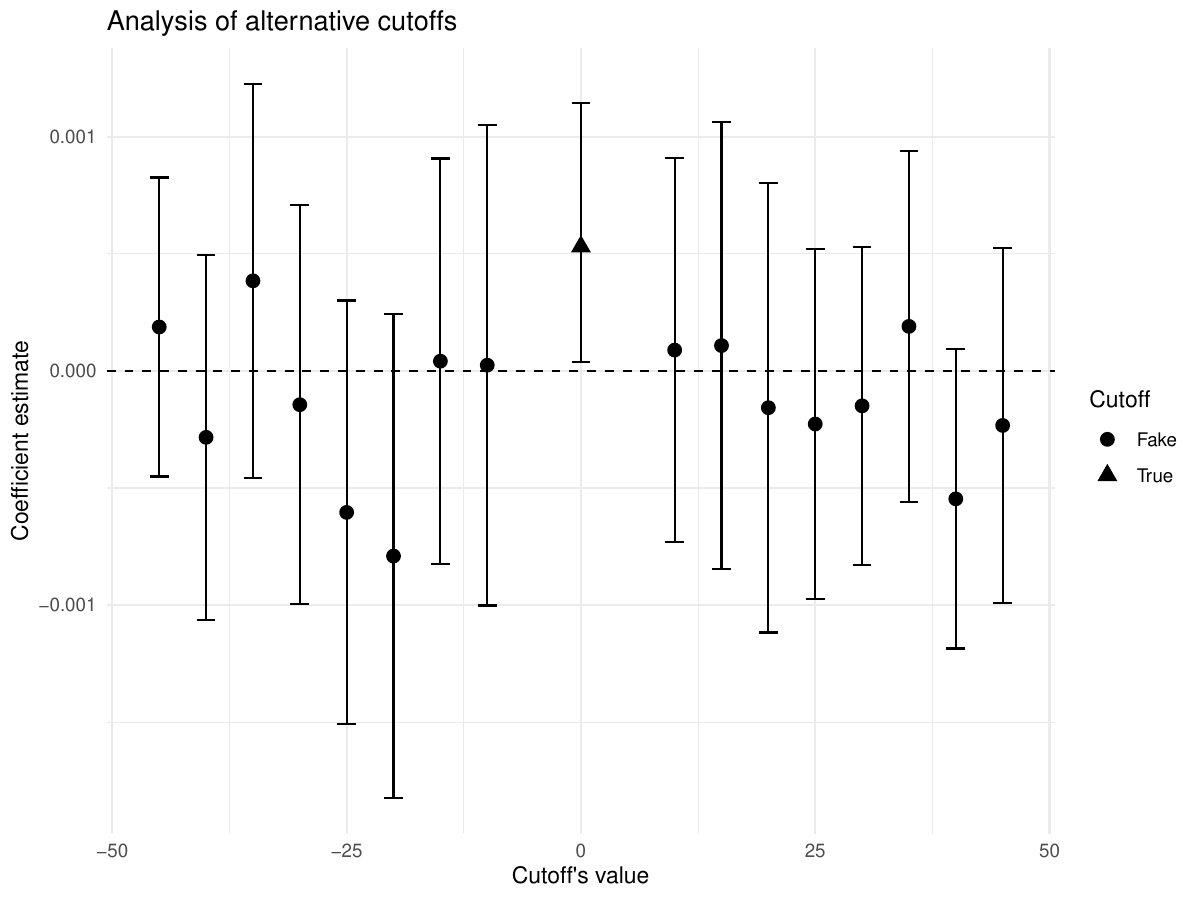}
    \caption{\textsc{Placebo test with fake threshold - RD} \\  The plot displays the coefficients and 90\% confidence intervals of the RD regressions, comparing the correct threshold (triangle) with fake thresholds (dots).}
    \label{fig:alternative}
\end{figure}

Finally, we perform a falsification check by examining the presence of a treatment effect at placebo cutoffs. This test provides further reassurance regarding the continuity of the regression functions for both treatment and control groups, confirming the absence of abrupt changes at the cutoff in the absence of the actual treatment. We implemented this by selecting values around the true cutoff and re-estimating the OPS procedure for each of them, treating control and treatment units separately \citep{Cattaneo2019}. The results are plotted in \hyperref[fig:alternative]{Figure~\ref*{fig:alternative}} with the inclusion of 90\% confidence intervals. No coefficient is significant, and this, together with the density test, provides strong support for the internal validity of the RDD estimates.

\FloatBarrier

\section{Conclusions}
\label{s:conclusions}


This paper provides a comprehensive analysis of the short- and medium-term impacts of hiring incentives for the conversion of fixed-term into open-ended contracts, also considering possible heterogeneity in impacts and spillover effects.
We exploit a 2018 age eligibility change in the Italian incentive schemes, which temporarily extended eligibility to workers under 35. This setting offers a unique opportunity to evaluate the effectiveness of hiring incentives among workers already employed on temporary contracts, in contrast to most existing studies that focus on workers who are either new or estranged from labor markets. It also isolates incentives targeted at new permanent hires from broader hiring schemes. 

Using rich administrative data from Tuscany, which allows us to track the career advancement of the working-age population, we employ a combination of regression discontinuity and difference-in-differences designs. Our results show that the incentives had an immediate positive impact on the conversion of temporary contracts into permanent ones, with no evidence of substitution against slightly older, ineligible cohorts. These short-term effects were sizable, suggesting that employers responded quickly to the temporary change in eligibility. However, we find that these positive effects were short-lived: within one year, the discontinuity vanishes, and no persistent differences emerge in terms of job stability, continued employment, or career progression. In practice, the incentives appear to have merely anticipated conversions that would have occurred in the absence of the policy.

Taken together, these findings suggest that while social contributions cuts can generate immediate boosts in permanent hiring, their medium-run effectiveness is limited when applied to workers already in employment relationships. For this group, employer decisions seem to be driven less by structural constraints and more by timing, raising doubts about the cost-effectiveness of temporary expansions in eligibility. Future research should continue to examine not only the conditions under which hiring incentives can create lasting employment gains, but also whether they are capable of doing so at all, particularly in labor markets where temporary contracts represent a structural entry point.

\newpage
\subsection*{Acknowledgements}
The authors would like to thank the Region of Tuscany for providing access to the data used in this study.

\subsection*{Declaration of Generative AI and AI-Assisted Technologies in the Manuscript Preparation Process}

During the preparation of this manuscript, the author(s) made use of AI-assisted technologies to enhance efficiency in editing and grammar checking. Following the use of these tools, the author(s) thoroughly reviewed and revised the content as necessary and assume full responsibility for the final version of the publishedarticle.

\bibliographystyle{myapalike}
\bibliography{Working_Paper_2.0.bib}

\clearpage
\appendix
\section*{Appendix}
\addcontentsline{toc}{section}{Appendix} 

\renewcommand{\thesection}{\Alph{section}}
\renewcommand{\thefigure}{\Alph{section}.\arabic{figure}}
\renewcommand{\thetable}{\Alph{section}.\arabic{table}}
\counterwithin{figure}{section}
\counterwithin{table}{section}

\section{Concurrent policies}
\label{s:ddignita}

The Decreto Dignità (Decree-Law No. 87 of 12 July 2018) was adopted by the Italian government and subsequently converted into law (Law No. 96 of 9 August 2018) after parliamentary scrutiny and amendments. It was initially approved in mid-July and became effective shortly thereafter.  At this stage, the new rules on fixed-term employment contracts applied primarily to newly stipulated contracts. Subsequently, through conversion into law (Law 9 August 2018, n. 96) and accompanying ministerial/circular clarifications, it was established that, from 31 October 2018 onward, the reform’s provisions would extend their reach to also cover renewals or extensions of contracts already in place as of 14 July 2018. Thus, from November 1, 2018, the new discipline was applied comprehensively.

The measure aimed to reduce job precariousness by tightening rules on fixed-term contracts and strengthening protections for permanent employment. 
Among the major modifications introduced: (i) the maximum duration of fixed-term contracts was reduced from 36 to 24 months, (ii) the maximum number of fixed-term contract renewals was cut from five to four, (iii) the requirement to report an explicit causal reason (“causale”) for the termination of a contract was extended to fixed-term contracts exceeding 12 months; (iv) social contributions labour costs were increased from 1,4\% to 1,9\%.

Several caveats apply, limiting the impact of this policy on our estimates. Firstly, it is essential to note that the new Decree-Law does not directly impact conversion per se and does not affect the 35-year-old age threshold that defines the group of eligible beneficiaries. The introduction of this policy can, then, at worst, moderate the effects of the incentives by increasing overall conversions.  Secondly, the policy only became effective in late 2018, meaning that only 36\% of contracts started after the initial partial implementation of the decree.

\begin{figure}
    \centering
    \includegraphics[width=0.7\linewidth]{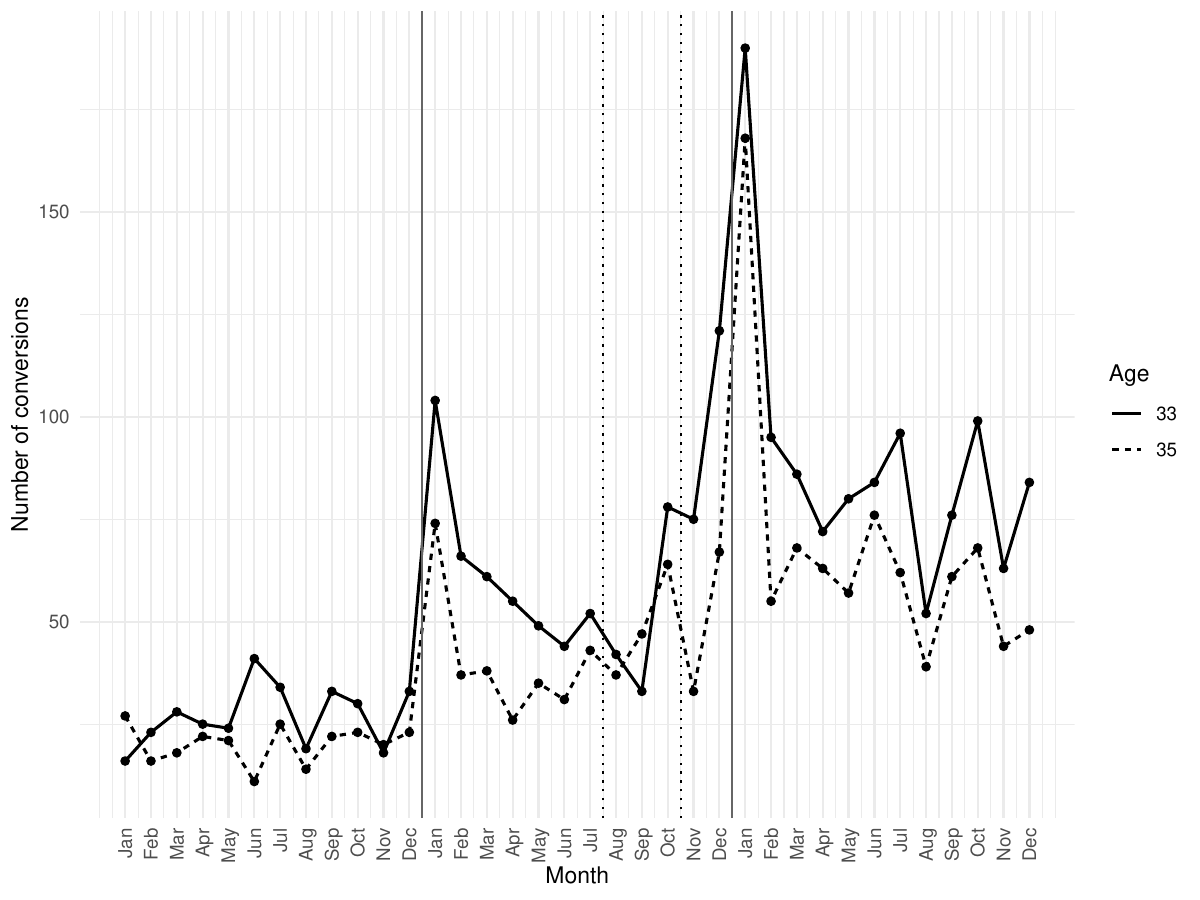}
    \caption{\textsc{Conversions by months from 2017 to 2019} \\  The plot displays the monthly conversions divided by people aged 33 and 35. The vertical solid lines indicate the boundaries between years, while the two dotted ones mark the partial (first) and full (second) implementation of the Decreto Dignità.}
    \label{fig:dd_plot}
\end{figure}

In light of these considerations, we can compare contract conversions before and after the introduction of the concurrent policy. \hyperref[fig:dd_plot]{Figure~\ref{fig:dd_plot}} shows that seasonal factors predominantly explain the observed dynamics, as the three years follow a comparable monthly pattern, though with a positive underlying trend. The main differences in conversion rates between the two age groups emerge in the early months of the year 2018, before the approval of the Degree. This pattern suggests that, once seasonality is properly controlled for, the short-term impact of the Decreto Dignità on the likelihood of contract conversion appears to be limited.
Regression results also seem to point in the same direction. In Table \ref{tab:DD} we report the estimates of the treatment interacted with a dummy variable equal to one for individuals born before August 1st. This variable allows us to examine potential heterogeneity between those who were simultaneously eligible for both reforms, the incentives and the Decreto Dignità, and other individuals. We report only the relevant coefficients, which indicate no meaningful difference between the two subgroups, suggesting a negligible effect on conversions of the more recent Decree.

Nonetheless, we cannot entirely rule out the possibility that, in the medium term, the Decreto Dignità may have either moderated or mitigated the observed incentive effect. The absence of interaction effects in the short term suggests that the results should also hold in the medium term. Hence, while some degree of moderation cannot be excluded, the identification strategy seems to remain valid, as does the internal validity of the analysis. 
\begin{table}[h!]
\centering
\title{\textbf{TABLE A.1: Interaction with Decreto Dignità}}
\begin{tabular}[t]{lc}
\toprule
  & Covariate Model\\
\midrule
Eligible & \num{-0.005}\\
 & \vphantom{8} (\num{0.005})\\
Born before August & \num{-0.006}\\
 & \vphantom{7} (\num{0.004})\\
Eligible × (t = 0) & \num{0.013}*\\
 & \vphantom{2} (\num{0.007})\\
Eligible × Born before August × (t = 0) & \num{0.002}\\
 & (\num{0.010})\\
\midrule
Num.Obs. & \num{113576}\\
Std.Errors & by: Firm\\
Covariates & \checkmark\\
\bottomrule
\multicolumn{2}{l}{\rule{0pt}{1em}Standard errors are in parentheses.}\\
\multicolumn{2}{l}{\rule{0pt}{1em}Significance levels: * p $<$ 0.1, ** p $<$ 0.05, *** p $<$ 0.01}\\
\end{tabular}
\caption{}
\label{tab:DD}
\end{table}

\section{Covariates distribution}
\label{Appendix:covariates}

\begin{table}[!ht]
\centering
\title{\textbf{TABLE B.1: Covariates balance}}
\resizebox{\textwidth}{!}{
\begin{tabular}[t]{lccccccc}
\toprule
  & Male & \makecell{Number of \\ renewals} & \makecell{Potential \\ duration} & Agricolture & Industry & Services & \makecell{Italian \\ citizenship}\\
\midrule
Eligible × (t = -4) & \num{-0.008} & \num{0.001} & \num{-0.139} & \num{0.003} & \num{-0.005} & \num{0.002} & \num{-0.032}\\
 & (\num{0.023}) & (\num{0.004}) & (\num{0.151}) & (\num{0.010}) & (\num{0.009}) & (\num{0.013}) & (\num{0.021})\\
Eligible × (t = -3) & \num{0.007} & \num{0.001} & \num{-0.292}** & \num{-0.006} & \num{-0.007} & \num{0.013} & \num{-0.032}\\
 & (\num{0.025}) & (\num{0.004}) & (\num{0.138}) & (\num{0.011}) & (\num{0.010}) & (\num{0.013}) & (\num{0.020})\\
Eligible × (t = -2) & \num{-0.018} & \num{0.000} & \num{-0.320}* & \num{-0.009} & \num{-0.013} & \num{0.022} & \num{-0.052}**\\
 & (\num{0.024}) & (\num{0.005}) & (\num{0.177}) & (\num{0.013}) & (\num{0.011}) & (\num{0.018}) & (\num{0.026})\\
Eligible × (t = 0) & \num{-0.028} & \num{-0.004} & \num{-0.216} & \num{0.007} & \num{-0.006} & \num{-0.001} & \num{0.022}\\
 & (\num{0.023}) & (\num{0.006}) & (\num{0.170}) & (\num{0.010}) & (\num{0.010}) & (\num{0.013}) & (\num{0.021})\\
Eligible × (t = 1) & \num{-0.037} & \num{-0.009} & \num{0.045} & \num{0.006} & \num{-0.007} & \num{0.001} & \num{-0.040}*\\
 & (\num{0.028}) & (\num{0.005}) & (\num{0.144}) & (\num{0.011}) & (\num{0.010}) & (\num{0.015}) & (\num{0.021})\\
\bottomrule
\footnotesize \\
\vspace{0.1cm}
Standard errors in parentheses. \\
Significance level: $^{*} p < 0.1$, $^{**} p < 0.05$, $^{***} p < 0.01$
\end{tabular}
}
\caption{}
\label{tab:balance}

\end{table}
\clearpage
\section{Difference in Differences}
\label{Appendix:did}

\begin{table}[!ht]
\title{\textbf{TABLE C.1: Difference in differences with spillover checks}}
\centering
\resizebox{0.90\textwidth}{!}{
\begin{tabular}{llll|l}
\toprule
\midrule
 & \makecell{Control\\ Age 37} & \makecell{Control\\ Age 38} & \makecell{Control\\ Age 39} & \makecell{Interaction with \\ share eligible$_{t-1}$}\\
\midrule
Eligible $\times$  (t=0) & \makecell{0.0108$^{**}$ \\ (0.0055)} & \makecell{0.0139$^{***}$ \\ (0.0054)} & \makecell{0.0120$^{**}$ \\ (0.0055)} & \makecell{0.017$^{**}$ \\ (0.007)} \\
Eligible $\times$ Share eligible$_{t-1}$ $\times$ (t=0)  &   &   &   & \makecell{-0.016 \\ (0.012)} \\
Eligible $\times$  (t=-2) & \makecell{0.0069 \\ (0.0048)} & \makecell{0.0057 \\ (0.0048)} & \makecell{-0.0029 \\ (0.0050)} & \makecell{0.011$^{*}$ \\ (0.007)} \\
Eligible $\times$  (t=-3) & \makecell{-0.0015 \\ (0.0048)} & \makecell{0.0036 \\ (0.0048)} & \makecell{0.0033 \\ (0.0049)} & \makecell{0.011$^{*}$ \\ (0.007)} \\
Eligible $\times$ (t=-4) & \makecell{-0.0001 \\ (0.0042)} & \makecell{0.0017 \\ (0.0043)} & \makecell{0.0027 \\ (0.0042)} &  \\
Eligible $\times$  (t=1) & \makecell{0.0080 \\ (0.0057)} & \makecell{0.0116$^{**}$ \\ (0.0057)} & \makecell{0.0094 \\ (0.0057)} & \makecell{0.017$^{**}$ \\ (0.007)} \\

Eligible $\times$ Share eligible$_{t-1}$ $\times$ (t=-3) &   &   &   & \makecell{-0.022$^{*}$ \\ (0.012)} \\
Eligible $\times$ Share eligible$_{t-1}$ $\times$ (t=-2) &   &   &   & \makecell{-0.022$^{**}$ \\ (0.011)} \\
Eligible $\times$ Share eligible$_{t-1}$ $\times$ (t=1)  &   &   &   & \makecell{-0.019 \\ (0.013)} \\
t=-2 & \makecell{0.0008 \\ (0.0035)} & \makecell{0.0024 \\ (0.0037)} & \makecell{0.0113$^{***}$ \\ (0.0040)} & \makecell{0.014$^{***}$ \\ (0.005)} \\
t=-3 & \makecell{0.0201$^{***}$ \\ (0.0040)} & \makecell{0.0154$^{***}$ \\ (0.0037)} & \makecell{0.0162$^{***}$ \\ (0.0039)} & \makecell{0.027$^{***}$ \\ (0.005)} \\
t=-4 & \makecell{0.0000 \\ (0.0033)} & \makecell{-0.0013 \\ (0.0033)} & \makecell{-0.0019 \\ (0.0033)} &  \\
t=0  & \makecell{0.0157$^{***}$ \\ (0.0039)} & \makecell{0.0127$^{***}$ \\ (0.0039)} & \makecell{0.0146$^{***}$ \\ (0.0039)} & \makecell{0.017$^{***}$ \\ (0.005)} \\
t=1  & \makecell{0.0335$^{***}$ \\ (0.0046)} & \makecell{0.0301$^{***}$ \\ (0.0044)} & \makecell{0.0323$^{***}$ \\ (0.0045)} & \makecell{0.031$^{***}$ \\ (0.005)} \\

Share eligible$_{t-1}$ $\times$ (t=-3) &   &   &   & \makecell{-0.028$^{***}$ \\ (0.009)} \\
Share eligible$_{t-1}$ $\times$ (t=-2) &   &   &   & \makecell{-0.026$^{***}$ \\ (0.008)} \\
Share eligible$_{t-1}$ $\times$ (t=0)  &   &   &   & \makecell{-0.016$^{**}$ \\ (0.008)} \\
Share eligible$_{t-1}$ $\times$ (t=1)  &   &   &   & \makecell{-0.006 \\ (0.009)} \\
Eligible $\times$ Share eligible$_{t-1}$ &   &   &   & \makecell{0.014$^{*}$ \\ (0.008)} \\
\midrule
Observations & 72157 & 70475 & 69468 & 62786 \\
\bottomrule
\end{tabular}
}
\footnotesize \\
\vspace{0.1cm}
Standard errors in parentheses. \\
Significance level: $^{*} p < 0.1$, $^{**} p < 0.05$, $^{***} p < 0.01$
\caption{}
\label{tab:control_spillover}
\end{table}

\FloatBarrier

\clearpage

\section{Regression in Discontinuity and Difference in Discontinuity}
\label{Appendix:RDspillover}

\begin{table}[!ht]
\title{\textbf{TABLE D.1: RDD and RD-DiD regressions}}
\centering
\resizebox{0.9\textwidth}{!}{
\begin{tabular}[t]{lllll}
\toprule
\midrule
 & RDD base & RDD covariates & RD-DiD base & RD-DiD covariates\\
\midrule
Eligible  × Share eligible & \makecell{-0.0003 \\ (0.0005)} & \makecell{-0.0002 \\ (0.0005)} & \makecell{-0.0000 \\ (0.0004)} & \makecell{-0.0000 \\ (0.0004)}\\
Eligible & \makecell{0.0002 \\ (0.0003)} & \makecell{0.0002 \\ (0.0003)} & \makecell{-0.0002 \\ (0.0002)} & \makecell{-0.0002 \\ (0.0002)}\\
Share eligible & \makecell{0.0005 \\ (0.0004)} & \makecell{0.0004 \\ (0.0004)} & \makecell{0.0002 \\ (0.0003)} & \makecell{0.0002 \\ (0.0003)}\\

Eligible × Share eligible × (t=0) &  &  & \makecell{-0.0002 \\ (0.0006)} & \makecell{-0.0002 \\ (0.0006)}\\
Eligible × Share eligible × (t=-2) &  &  & \makecell{0.0000 \\ (0.0005)} & \makecell{0.0000 \\ (0.0005)}\\
Eligible × Share eligible × (t=-3) &  &  & \makecell{0.0005 \\ (0.0006)} & \makecell{0.0004 \\ (0.0006)}\\
Eligible  × Share eligible × (t=-4) &  &  & \makecell{0.0004 \\ (0.0005)} & \makecell{0.0004 \\ (0.0005)}\\
Eligible  × Share eligible × (t=1) &  &  & \makecell{-0.0014* \\ (0.0008)} & \makecell{-0.0014 \\ (0.0008)}\\
Eligible × (t=-2) &  &  & \makecell{0.0000 \\ (0.0002)} & \makecell{0.0000 \\ (0.0002)}\\
Eligible × (t=-3) &  &  & \makecell{-0.0002 \\ (0.0004)} & \makecell{-0.0002 \\ (0.0004)}\\
Eligible × (t=-4) &  &  & \makecell{0.0001 \\ (0.0002)} & \makecell{0.0001 \\ (0.0002)}\\
Eligible × (t=0) &  &  & \makecell{0.0004 \\ (0.0003)} & \makecell{0.0004 \\ (0.0003)}\\
Eligible × (t=1) &  &  & \makecell{0.0014** \\ (0.0006)} & \makecell{0.0014** \\ (0.0006)}\\
Share eligible × (t=-2) &  &  & \makecell{-0.0001 \\ (0.0004)} & \makecell{-0.0001 \\ (0.0004)}\\
Share eligible × (t=-3) &  &  & \makecell{-0.0004 \\ (0.0005)} & \makecell{-0.0004 \\ (0.0005)}\\
Share eligible × (t=-4) &  &  & \makecell{-0.0003 \\ (0.0003)} & \makecell{-0.0003 \\ (0.0003)}\\
Share eligible × (t=0) &  &  & \makecell{0.0002 \\ (0.0005)} & \makecell{0.0002 \\ (0.0005)}\\
Share eligible × (t=1) &  &  & \makecell{0.0003 \\ (0.0005)} & \makecell{0.0003 \\ (0.0005)}\\
\midrule
Observations & 64432 & 64432 & 380629 & 380629\\
Covariates & & \checkmark & & \checkmark\\
\bottomrule
\end{tabular}
}
\footnotesize\\
\vspace{0.1cm}
All the standard errors are clustered at workplace and individual level\\
Significance level: $^{*} p < 0.1$, $^{**} p < 0.05$, $^{***} p < 0.01$
\caption{}
\label{tab:rdd_rdddid}
\end{table}

\begin{figure}[!ht]
\centering
\includegraphics[width=0.95\linewidth]{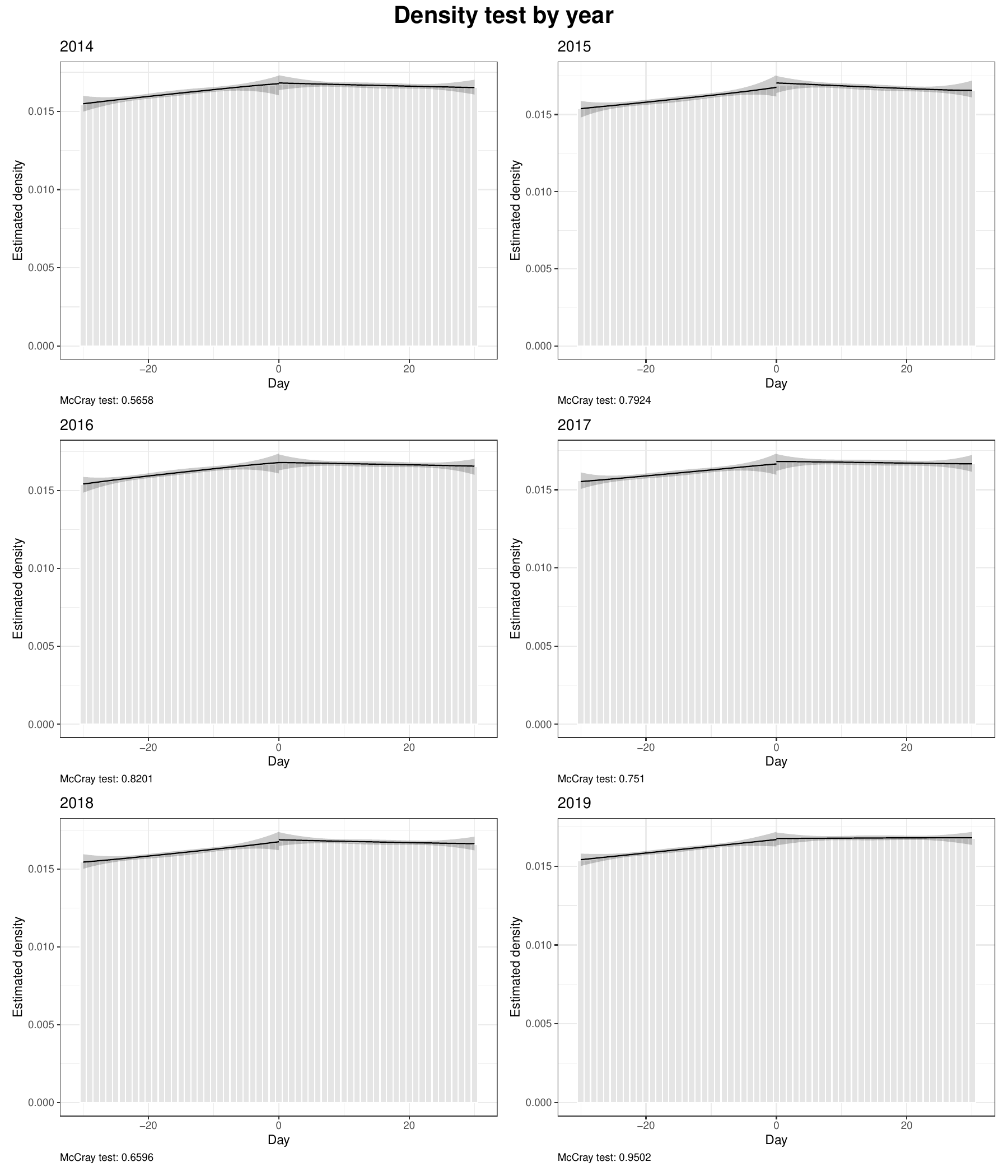}
\caption{\textsc{Density test for the RD-DiD} \\ The plot shows the distribution of the observations by each observation year, with 95\% confidence interval. Below the plots, the corresponding McCray test is also reported.}
\label{fig:figDENSITY}
\end{figure}
\FloatBarrier

\clearpage
\section{Medium run effects}
\label{Appendix:medium-run}
\begin{table}[!ht]
\centering
\title{\textbf{TABLE E.1: DiD medium-run outcomes two years after}}
\begin{tabular}[t]{llllll}
\toprule
\midrule
  & \makecell{Same \\ contract} & \makecell{Permanent \\ contract} & \makecell{Same \\ firm} & \makecell{Same \\ sector} & Days worked\\
\midrule
Eligible $\times$ (t=0)  & \makecell{-0.004 \\ (0.009)} & \makecell{0.012 \\ (0.015)} & \makecell{0.013 \\ (0.014)} & \makecell{0.014 \\ (0.014)} & \makecell{0.012 \\ (0.012)}\\
Eligible $\times$ (t=-4) & \makecell{-0.004 \\ (0.007)} & \makecell{-0.010 \\ (0.015)} & \makecell{-0.010 \\ (0.015)} & \makecell{-0.005 \\ (0.014)} & \makecell{-0.001 \\ (0.012)}\\
Eligible $\times$ (t=-3) & \makecell{-0.008 \\ (0.008)} & \makecell{-0.010 \\ (0.015)} & \makecell{-0.004 \\ (0.015)} & \makecell{-0.006 \\ (0.015)} & \makecell{-0.002 \\ (0.013)}\\
Eligible $\times$ (t=-2) & \makecell{-0.010 \\ (0.008)} & \makecell{-0.028* \\ (0.015)} & \makecell{-0.027* \\ (0.015)} & \makecell{-0.030** \\ (0.014)} & \makecell{-0.018 \\ (0.012)}\\

t=-4 & \makecell{-0.043*** \\ (0.006)} & \makecell{-0.010 \\ (0.012)} & \makecell{-0.009 \\ (0.011)} & \makecell{-0.019* \\ (0.011)} & \makecell{-0.009 \\ (0.009)}\\
t=-3 & \makecell{-0.029*** \\ (0.006)} & \makecell{-0.001 \\ (0.011)} & \makecell{-0.008 \\ (0.011)} & \makecell{-0.019* \\ (0.011)} & \makecell{-0.021** \\ (0.009)}\\
t=-2 & \makecell{-0.019** \\ (0.006)}  & \makecell{-0.030*** \\ (0.011)} & \makecell{-0.033*** \\ (0.011)} & \makecell{-0.043*** \\ (0.011)} & \makecell{-0.040*** \\ (0.009)}\\
t=0 & \makecell{0.012 \\ (0.006)}     & \makecell{-0.007 \\ (0.011)} & \makecell{-0.005 \\ (0.010)} & \makecell{0.001 \\ (0.010)}  & \makecell{0.014* \\ (0.008)}\\
Eligible     & \makecell{0.013* \\ (0.006)}    & \makecell{0.020* \\ (0.011)} & \makecell{0.018* \\ (0.010)} & \makecell{0.021** \\ (0.010)} & \makecell{0.014 \\ (0.009)}\\

\midrule
Num. Obs. & 63646 & 35683& 35683&35683&35683\\
\bottomrule
\end{tabular}
\footnotesize\\
\vspace{0.1cm}
All the standard errors are clustered at workplace level\\
significance level: $^{*} p < 0.1$, $^{**} p < 0.05$, $^{***} p < 0.01$
\caption{}
\label{tab:med_2}
\end{table}

\begin{table}[!ht]
\centering
\title{\textbf{TABLE E.2: DiD medium-run outcomes three years after}}
\begin{tabular}[t]{llllll}
\toprule
\midrule
  & \makecell{Same \\ contract} & \makecell{Permanent \\ contract} & \makecell{Same \\ firm} & \makecell{Same \\ sector} & Days worked\\
\midrule
Eligible $\times$ (t=0)  & \makecell{0.002 \\ (0.008)} & \makecell{0.013 \\ (0.015)} & \makecell{0.014 \\ (0.014)} & \makecell{0.015 \\ (0.014)} & \makecell{0.007 \\ (0.011)}\\
Eligible $\times$ (t=-4) & \makecell{-0.001 \\ (0.006)} & \makecell{0.000 \\ (0.015)} & \makecell{0.000 \\ (0.015)} & \makecell{0.004 \\ (0.014)} & \makecell{0.008 \\ (0.011)}\\
Eligible $\times$ (t=-3) & \makecell{-0.000 \\ (0.007)} & \makecell{0.004 \\ (0.016)} & \makecell{0.008 \\ (0.015)} & \makecell{0.006 \\ (0.015)} & \makecell{0.003 \\ (0.012)}\\
Eligible $\times$ (t=-2) & \makecell{-0.004 \\ (0.007)} & \makecell{-0.024 \\ (0.015)} & \makecell{-0.023 \\ (0.015)} & \makecell{-0.029* \\ (0.015)} & \makecell{-0.014 \\ (0.012)}\\

t=-4 & \makecell{-0.036*** \\ (0.005)} & \makecell{-0.022* \\ (0.012)} & \makecell{-0.026** \\ (0.011)} & \makecell{-0.044*** \\ (0.011)} & \makecell{-0.045*** \\ (0.009)}\\
t=-3 & \makecell{-0.026*** \\ (0.005)} & \makecell{-0.003 \\ (0.012)} & \makecell{-0.014 \\ (0.011)} & \makecell{-0.031*** \\ (0.011)} & \makecell{-0.038*** \\ (0.009)}\\
t=-2 & \makecell{-0.019*** \\ (0.005)} & \makecell{-0.004 \\ (0.011)} & \makecell{-0.007 \\ (0.011)} & \makecell{-0.022** \\ (0.011)} & \makecell{-0.033*** \\ (0.009)}\\
t=0  & \makecell{0.012* \\ (0.005)}    & \makecell{0.003 \\ (0.011)} & \makecell{-0.003 \\ (0.011)} & \makecell{-0.003 \\ (0.010)} & \makecell{-0.004 \\ (0.008)}\\
Eligible     & \makecell{0.008 \\ (0.005)}     & \makecell{0.016 \\ (0.011)} & \makecell{0.016 \\ (0.010)} & \makecell{0.018* \\ (0.010)} & \makecell{0.013 \\ (0.008)}\\

\midrule
Num. Obs. & 63646 & 35683& 35683&35683&35683\\
\bottomrule
\end{tabular}
\footnotesize\\
\vspace{0.1cm}
All the standard errors are clustered at workplace level\\
significance level: $^{*} p < 0.1$, $^{**} p < 0.05$, $^{***} p < 0.01$
\caption{}
\label{tab:med_3}

\end{table}
\FloatBarrier

\clearpage
\section{Heterogeneity}
\subsection{Short run}
\label{Appendix:heterogeneity}
\begin{figure}[htbp]
    \caption{ }
  \centering
  \begin{subfigure}{0.48\textwidth}
    \includegraphics[width=\linewidth]{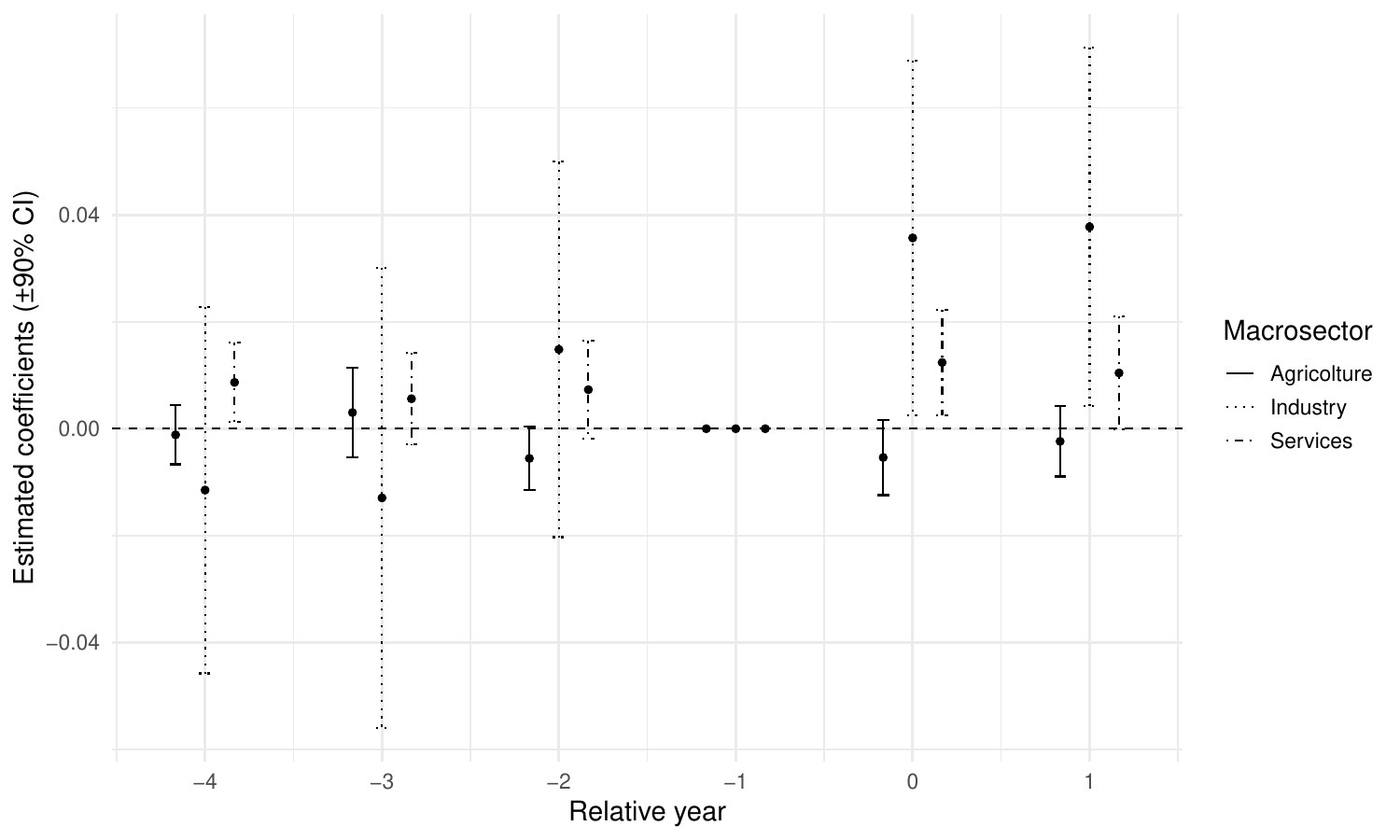}
    \caption*{Panel a - Macrosector}
  \end{subfigure}
  \hfill
  \begin{subfigure}{0.48\textwidth}
    \includegraphics[width=\linewidth]{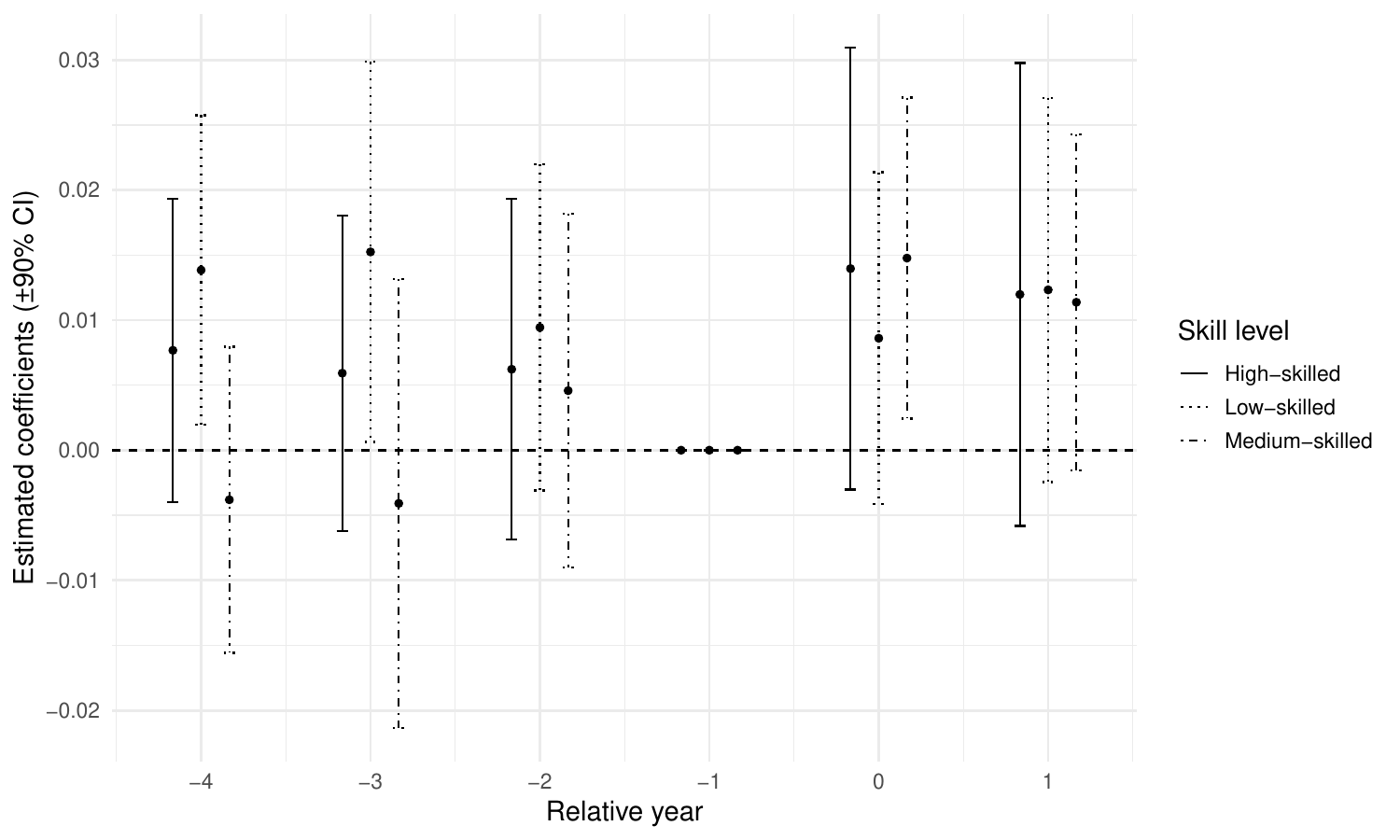}
    \caption*{Panel b - Skill Level}
  \end{subfigure}

  \vspace{1em}

  \begin{subfigure}{0.48\textwidth}
    \includegraphics[width=\linewidth]{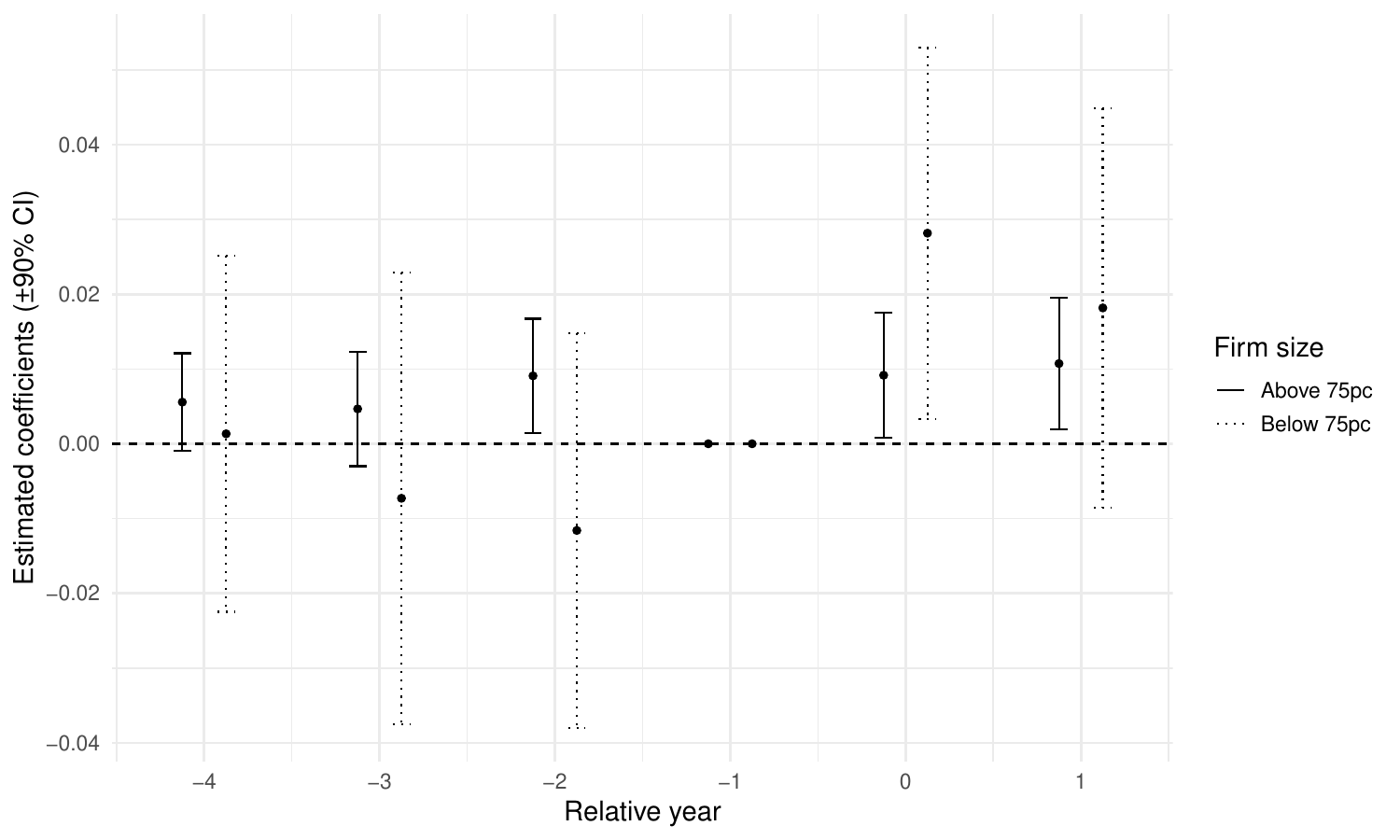}
    \caption*{Panel c - Firm Size}
  \end{subfigure}
  \hfill
  \begin{subfigure}{0.48\textwidth}
    \includegraphics[width=\linewidth]{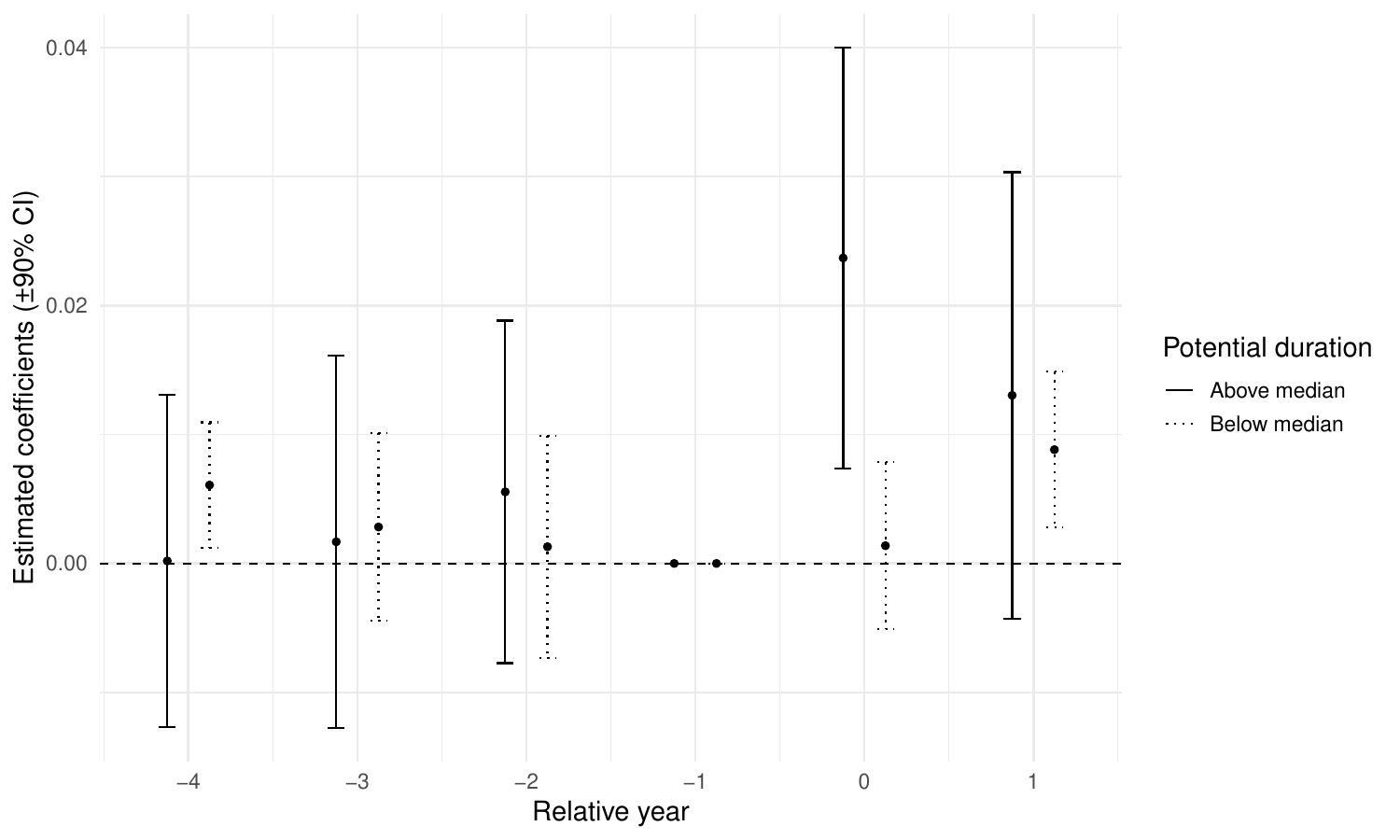}
    \caption*{Panel d - Potential Duration}
  \end{subfigure}

  \vspace{1em}

  \begin{subfigure}{0.48\textwidth}
    \includegraphics[width=\linewidth]{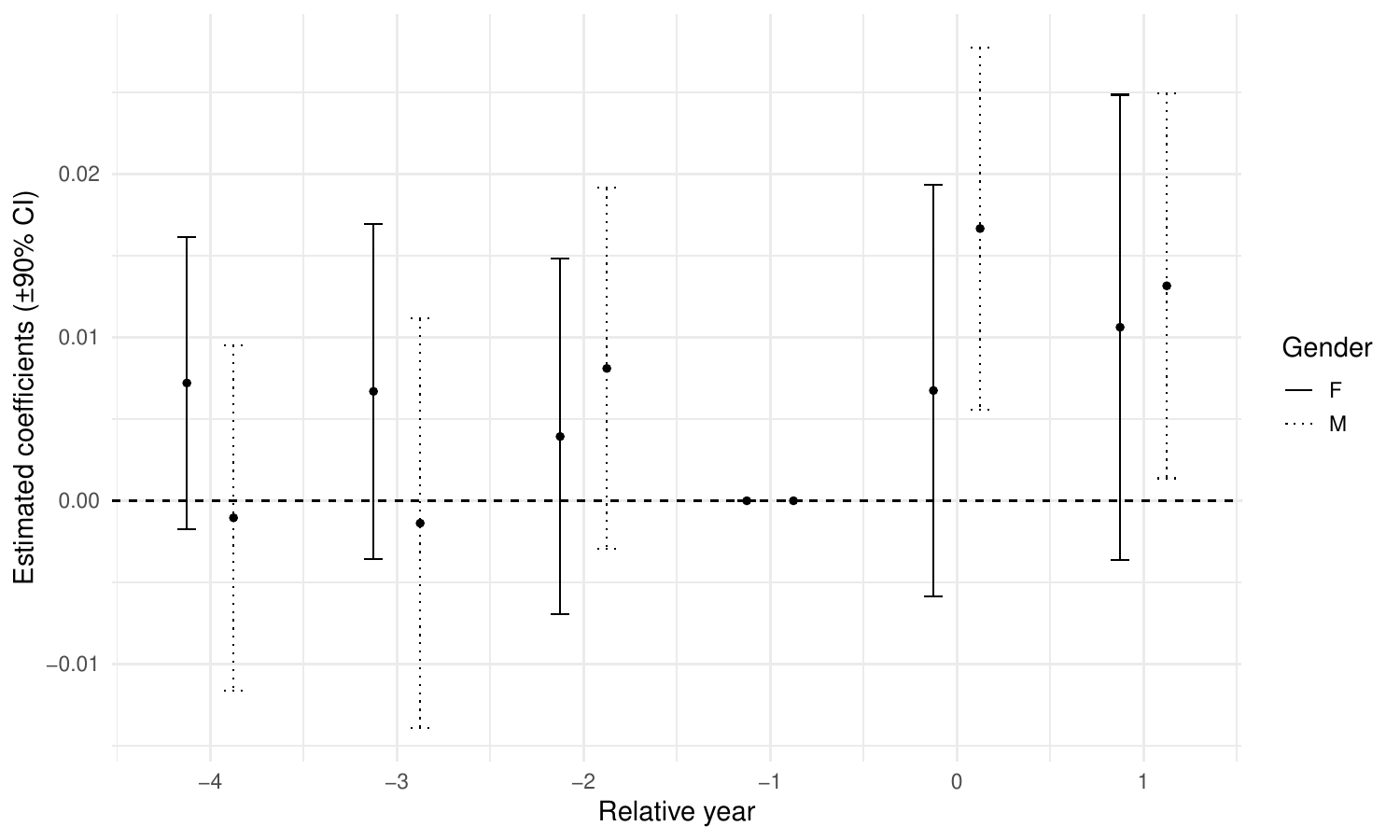}
\caption*{Panel e - Gender}
  \end{subfigure}
  \hfill
  \begin{subfigure}{0.48\textwidth}
  \end{subfigure}
  \label{fig:griglia}
  \end{figure}

\clearpage
\subsection{Medium run - 1 year forward}
\begin{figure}[h!]
    \caption{ }
    \label{fig:het2}
    \centering
    \includegraphics[width=0.925\linewidth]{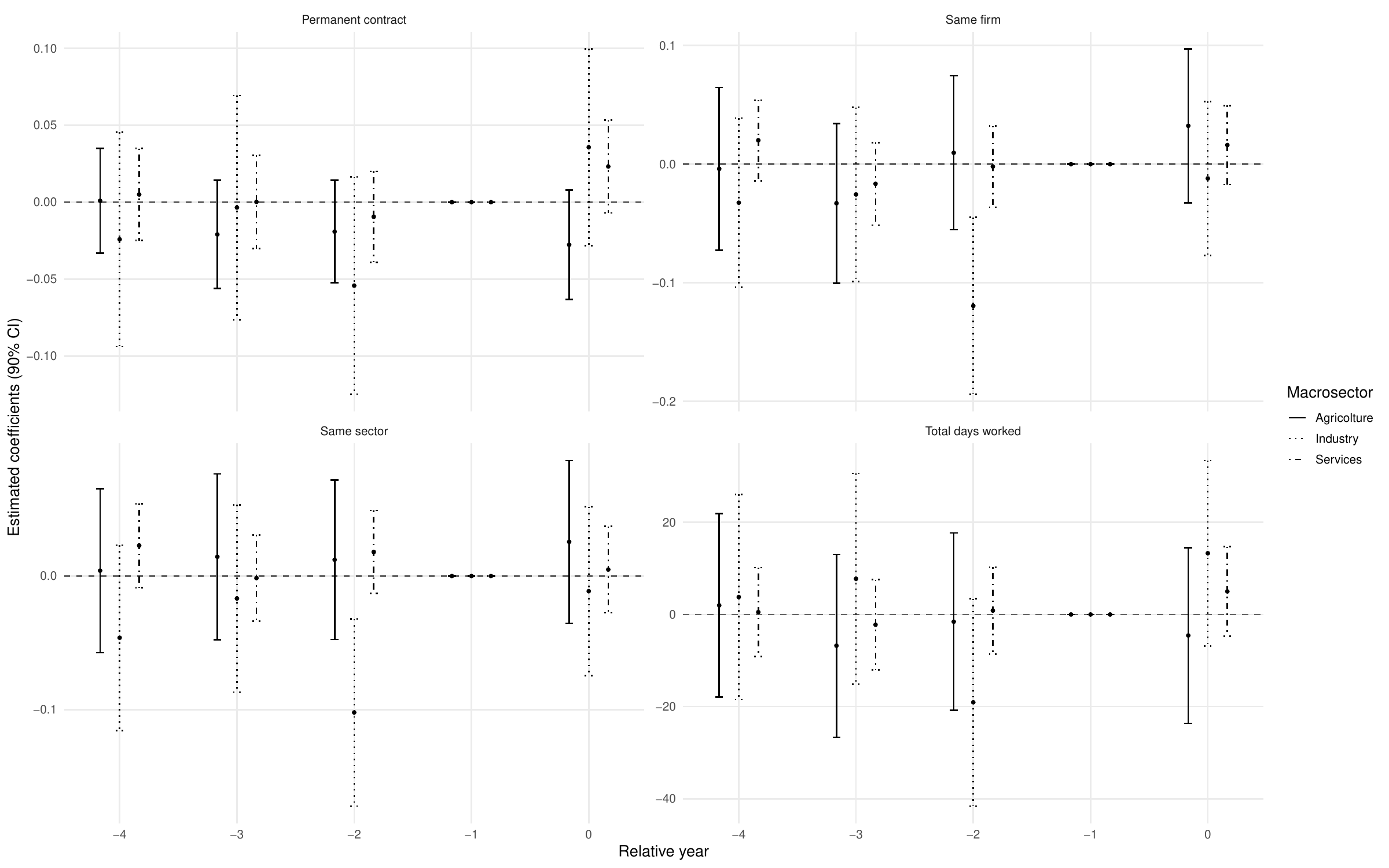}\\
        \caption*{Panel a - Macrosector}
    \includegraphics[width=0.925\linewidth]{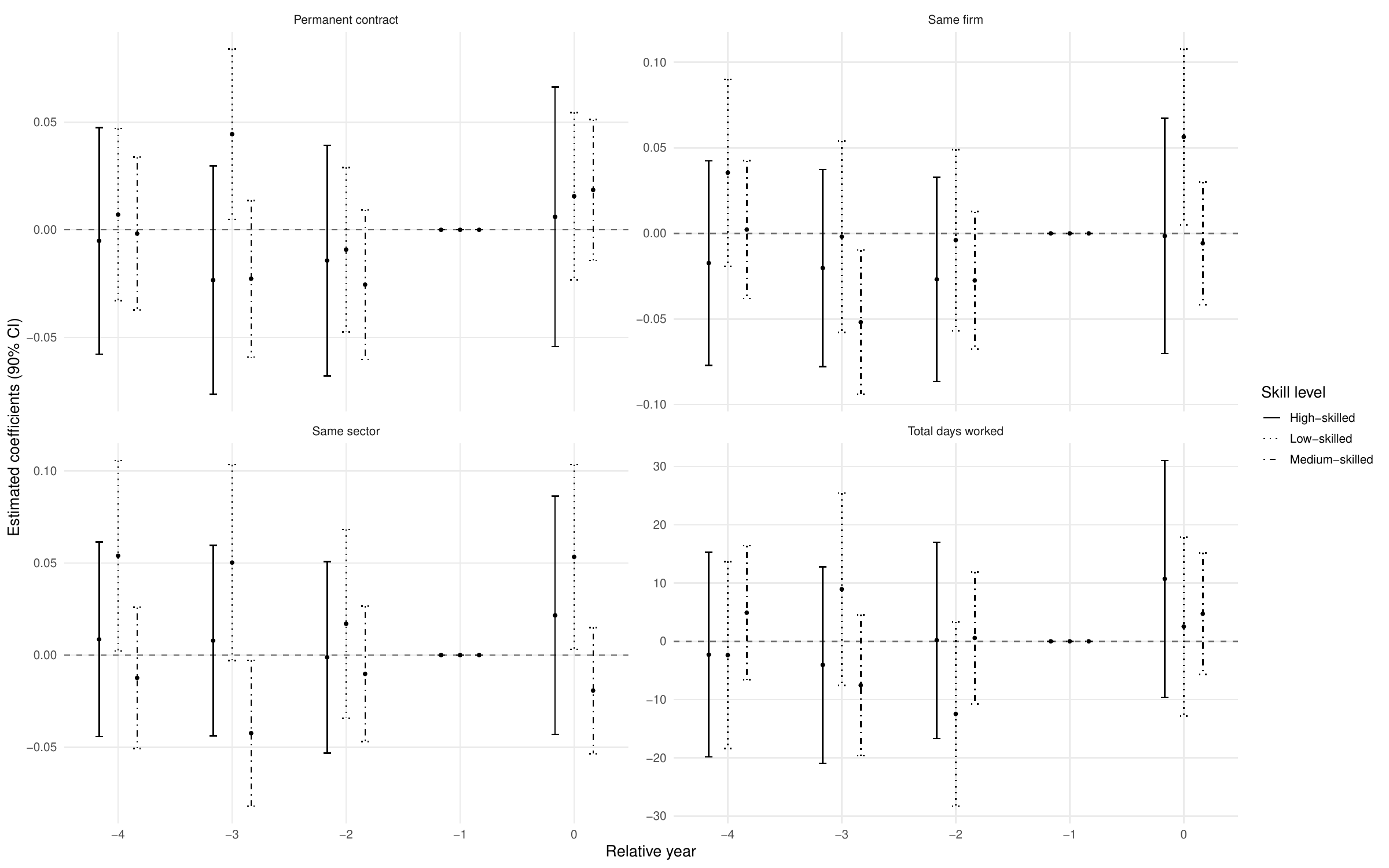}
        \caption*{Panel b - Skill Level}
\end{figure}
\clearpage

\begin{figure}[h!]
\centering
    \includegraphics[width=0.925\linewidth]{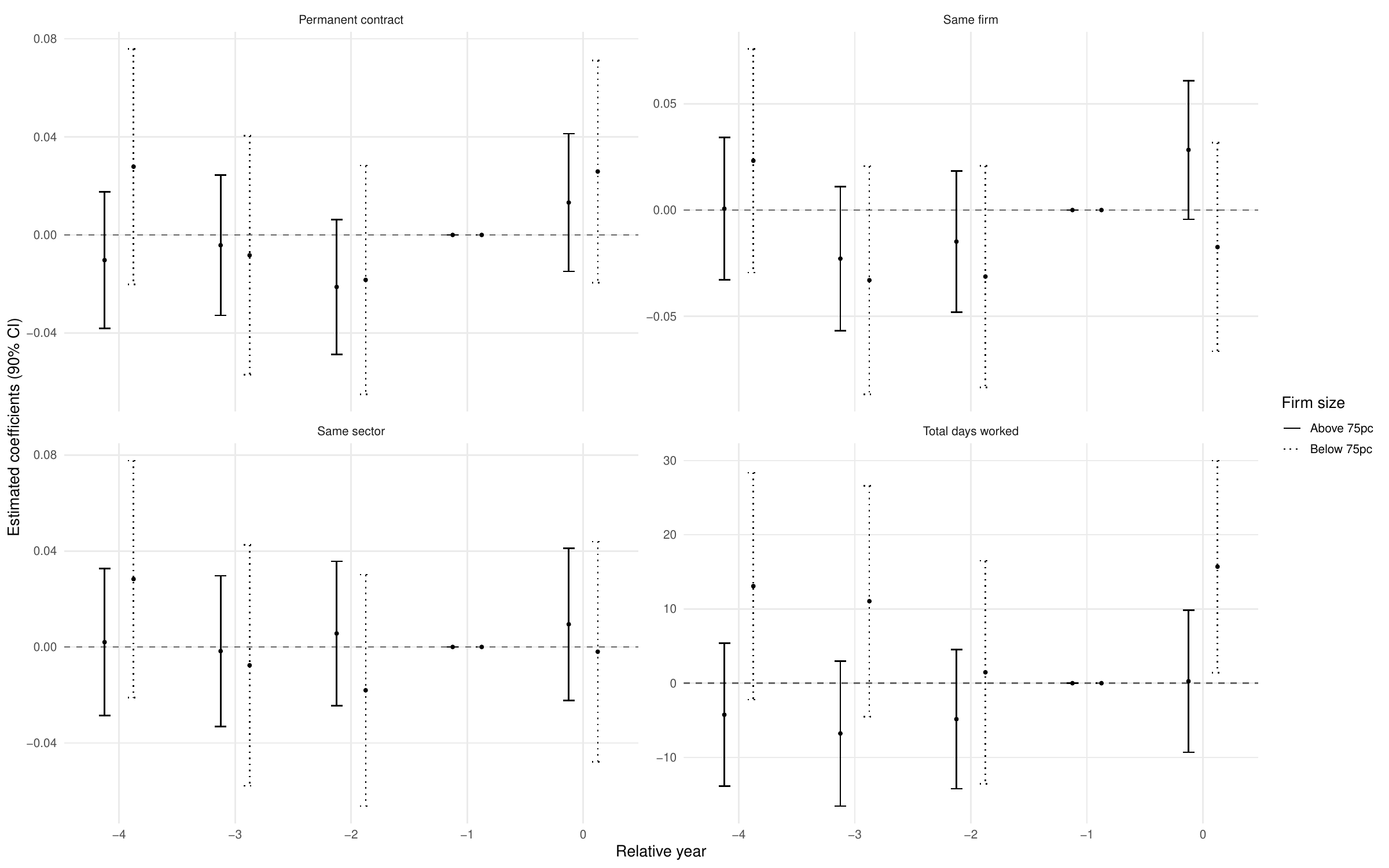}
        \caption*{Panel c - Firm Size}
\centering
    \includegraphics[width=0.925\linewidth]{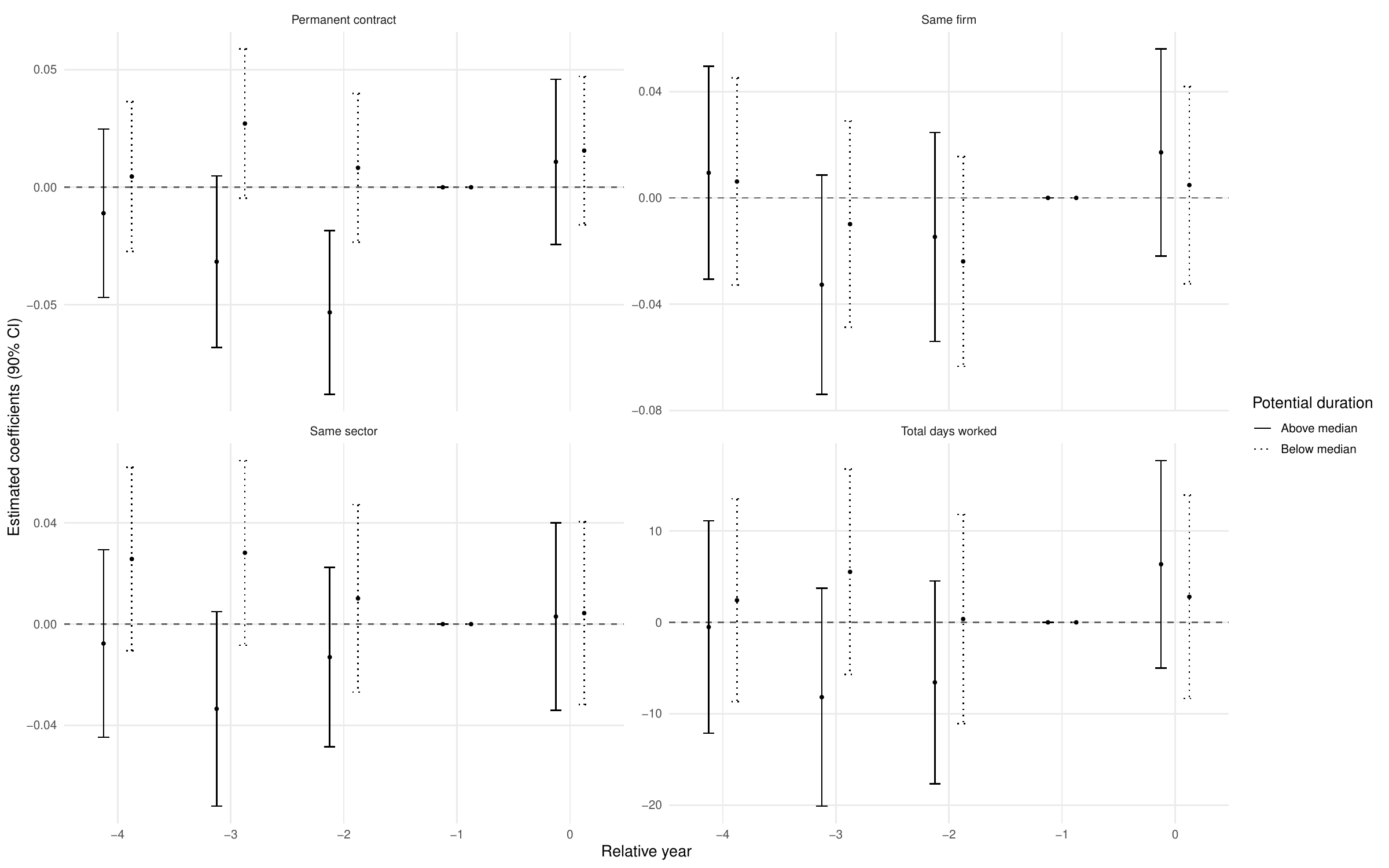}
        \caption*{Panel d - Potential Duration}
        \end{figure}
\clearpage

\begin{figure}[h!]
\centering
    \includegraphics[width=0.925\linewidth]{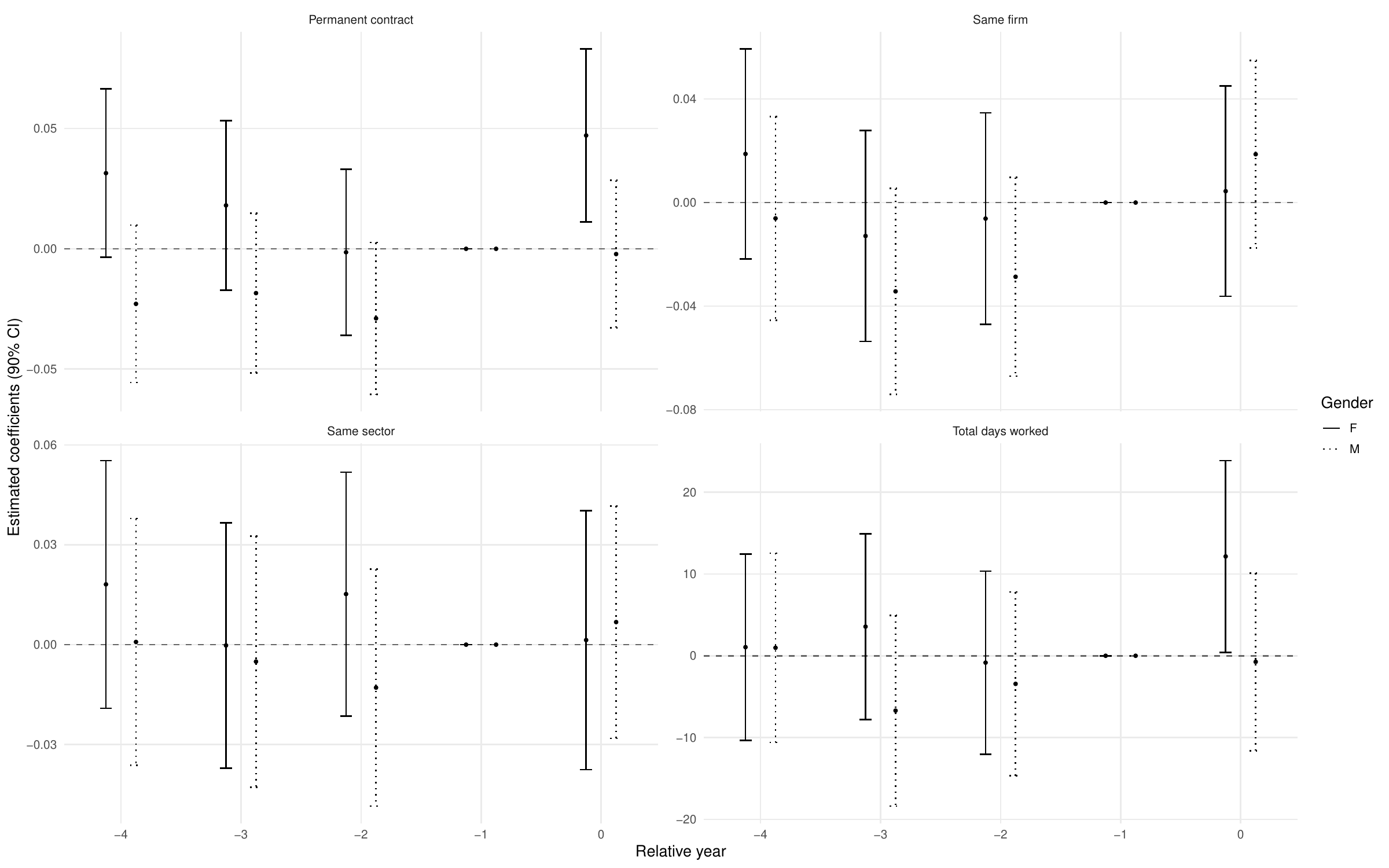}\\
        \caption*{Panel e - Gender}
\end{figure}
\clearpage

\subsection{Medium run - 4 year forward}
\begin{figure}[h!]
    \caption{ }
    \label{fig:het3}
    \centering
    \includegraphics[width=0.925\linewidth]{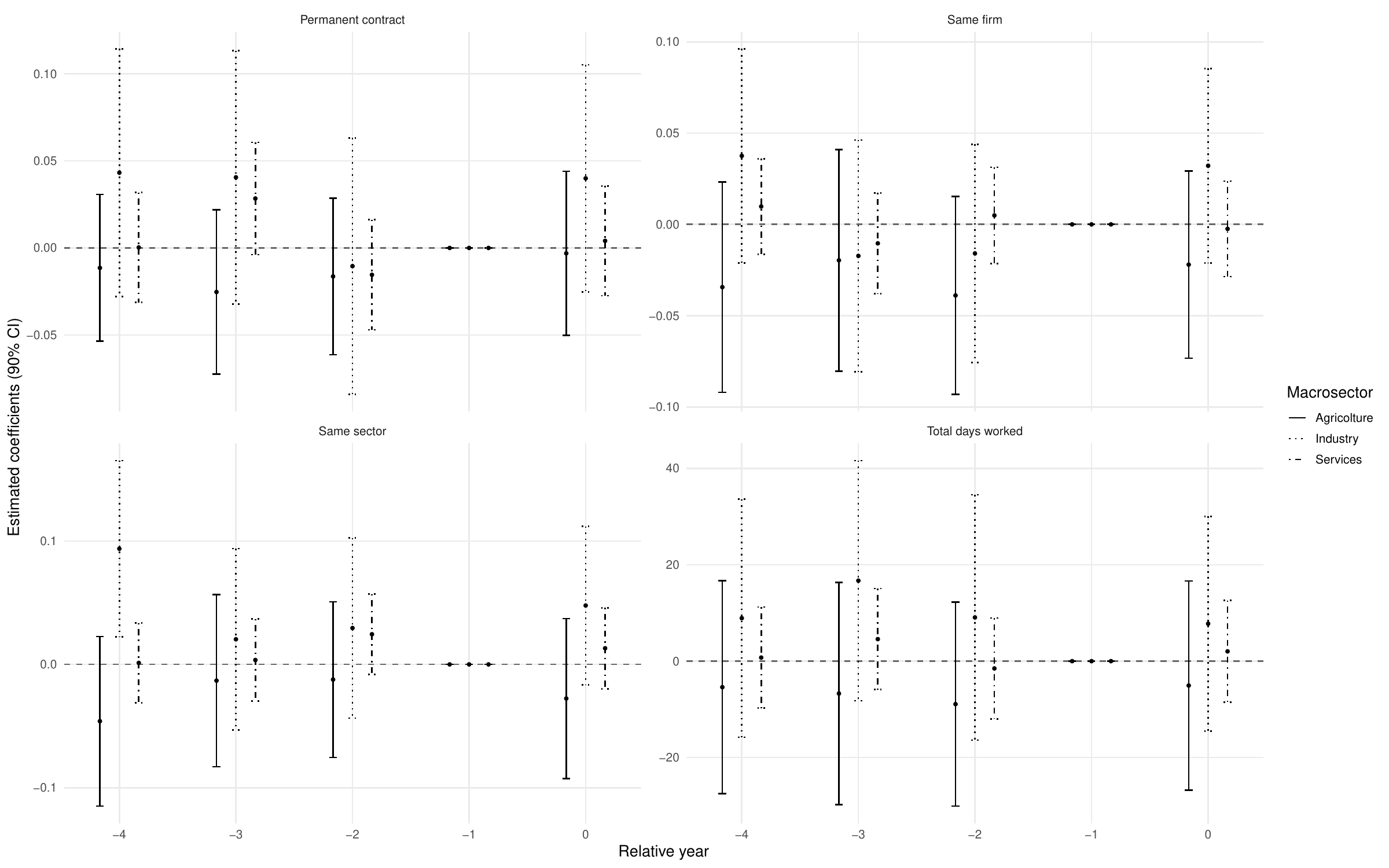}\\
        \caption*{Panel a - Macrosector}
    \includegraphics[width=0.925\linewidth]{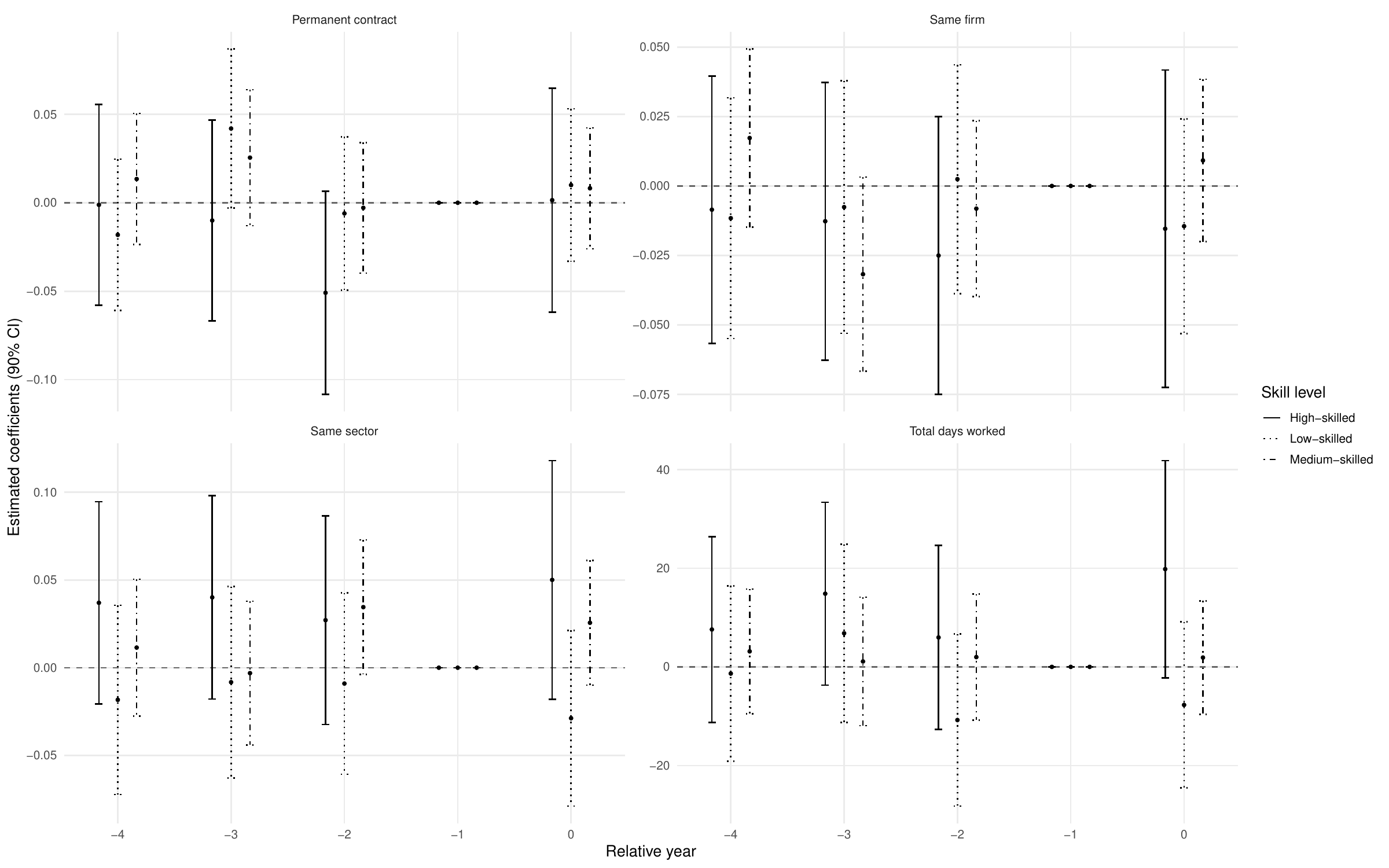}
        \caption*{Panel b - Skill Level}
\end{figure}
\clearpage

\begin{figure}[h!]
\centering
    \includegraphics[width=0.925\linewidth]{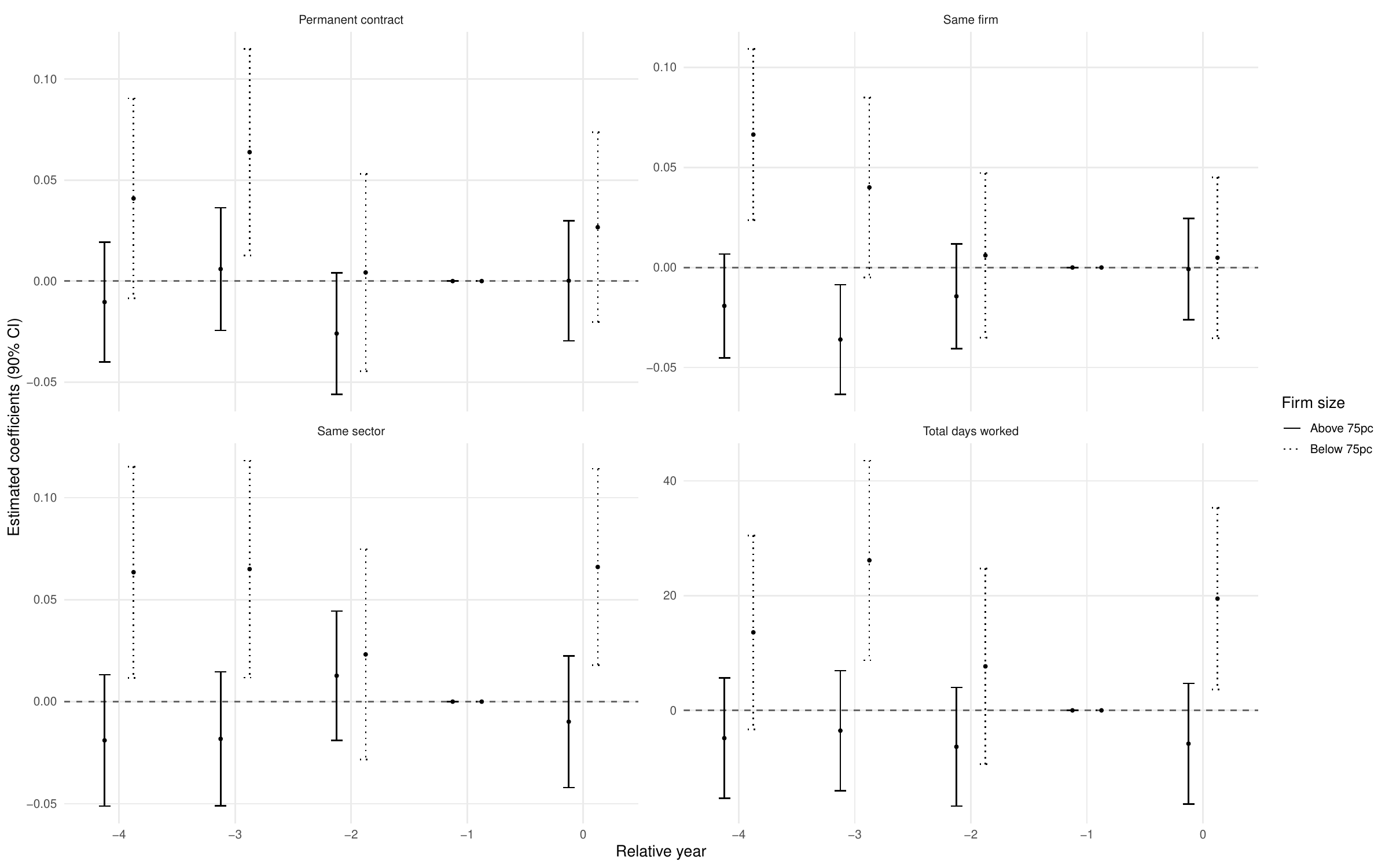}
        \caption*{Panel c - Firm Size}
\centering
    \includegraphics[width=0.95\linewidth]{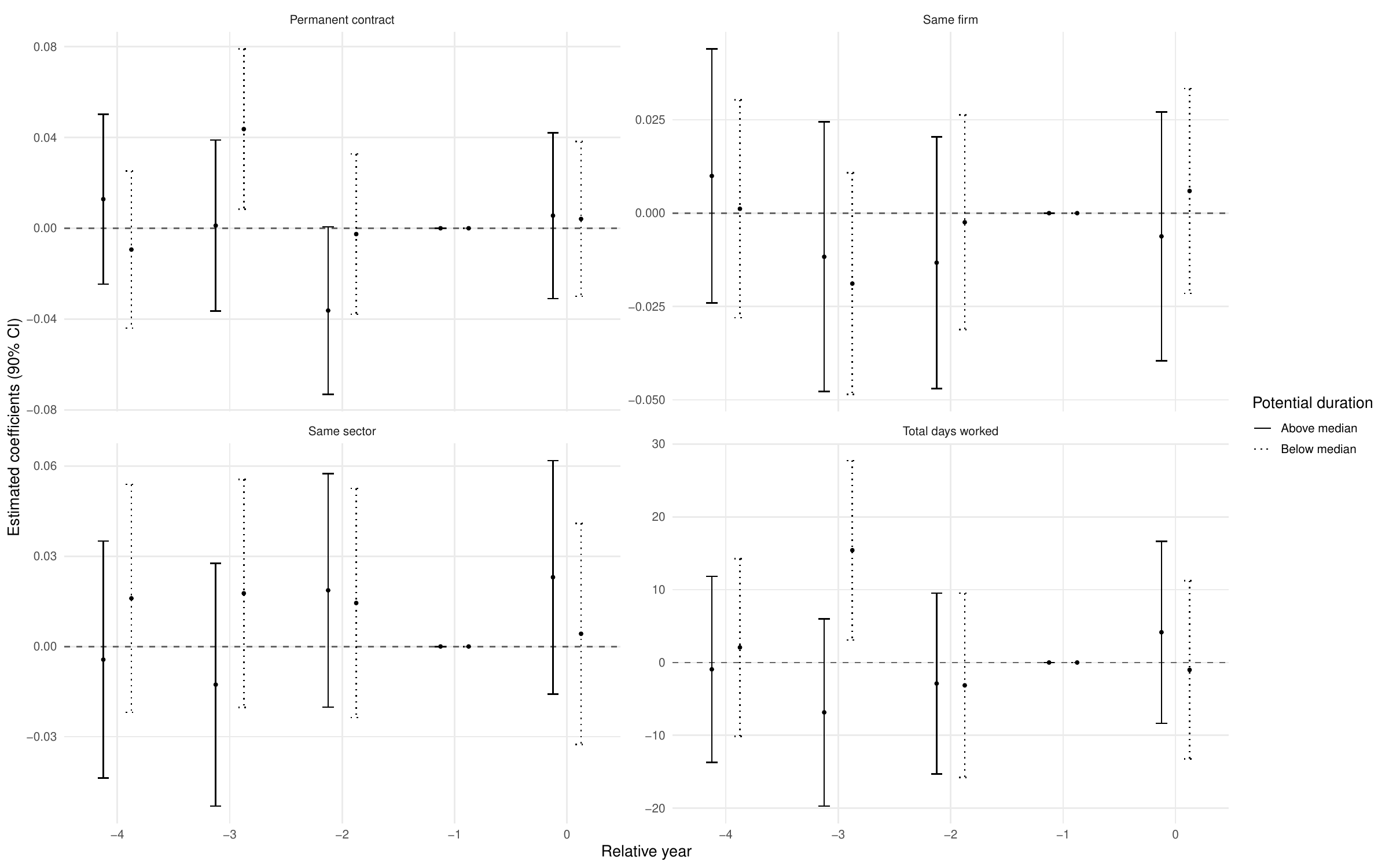} \caption*{Panel d - Potential Duration}
    \end{figure}
\clearpage

\begin{figure}[h!]
\centering
    \includegraphics[width=0.925\linewidth]{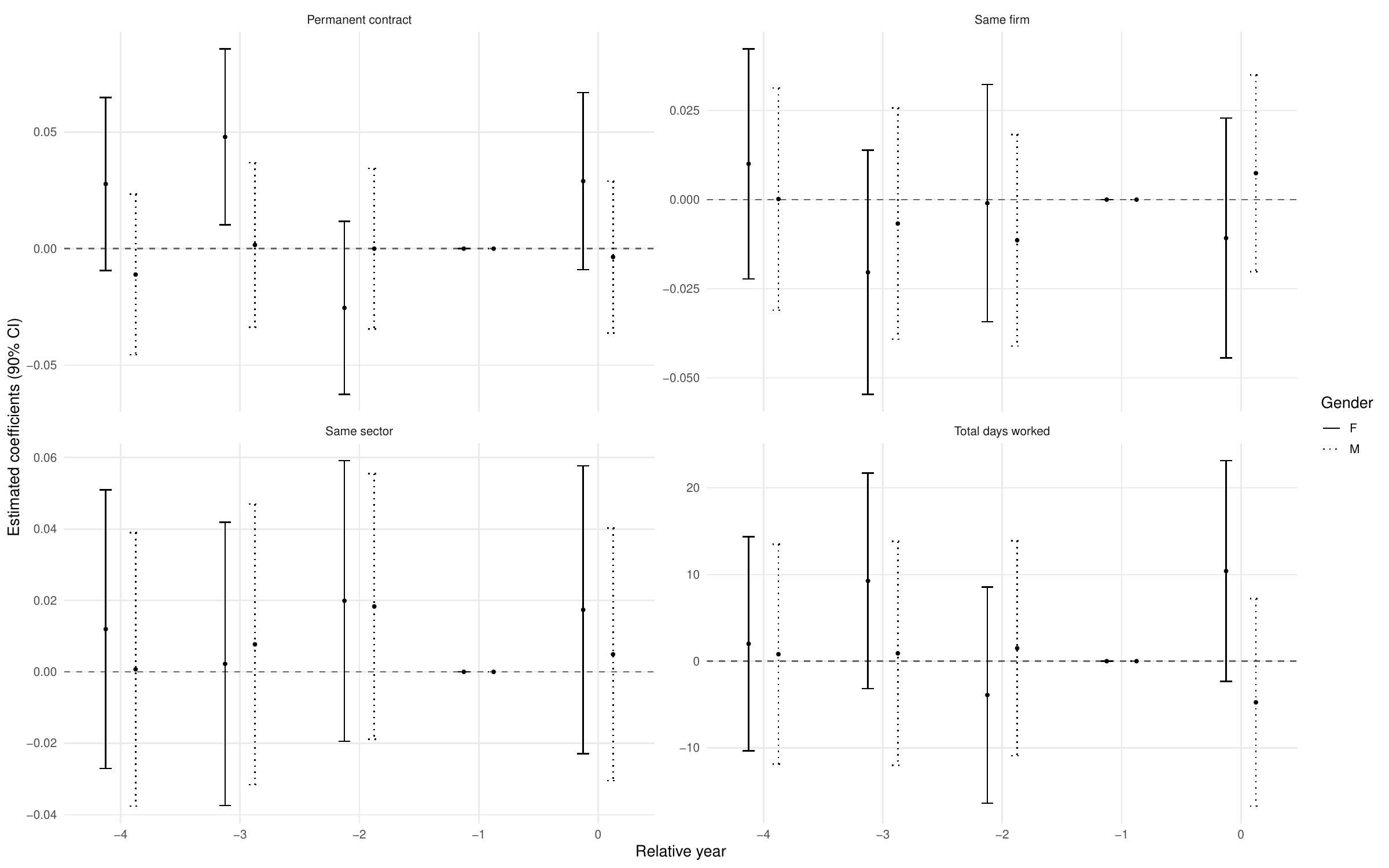}\\
        \caption*{Panel e - Gender}
\end{figure}
\end{document}